\documentclass[useAMS,usenatbib]{mn2e}

\usepackage[pdftex,pdfpagemode={UseOutlines},bookmarks,bookmarksopen,colorlinks,linkcolor={blue},citecolor={green},urlcolor={red}]{hyperref}
\usepackage{journals_macros}
\usepackage{graphicx}
\usepackage{natbib}
\usepackage{xspace}
\usepackage{amsmath}

\usepackage{times}

\def\rpe{r_{\perp}}
\def\rpa{r_{\parallel}}
\def\xisp{\xi(\rpe,\rpa)}

\def\xidd{\xi_{\rmn{\delta\delta}}}
\def\xidt{\xi_{\rmn{\delta\theta}}}
\def\xitt{\xi_{\rmn{\theta\theta}}}
\def\Pdd{P_{\rmn{\delta\delta}}}
\def\Pdt{P_{\rmn{\delta\theta}}}
\def\Ptt{P_{\rmn{\theta\theta}}}
\def\mhmpc{{\rm h}^{-1}~\rm{Mpc}}
\def\mkms{\,{\rm km \cdot s^{-1}}}

\newcommand{\aexp}{{\rm A}{\sc-exp}\xspace}

\newcommand{\bexp}{{\rm B}{\sc-exp}\xspace}

\newcommand{\cexp}{{\rm C}{\sc-exp}\xspace}
\newcommand{\cgauss}{{\rm C}{\sc-gauss}\xspace}
\newcommand{\hmpc}{$\,{\rm h}^{-1}$ Mpc\xspace}

\title[Modelling non-linear redshift-space distortions] {Modelling
  non-linear redshift-space distortions in the galaxy clustering
  pattern: systematic errors on the growth rate parameter}

\author[S. de la Torre \& L. Guzzo]
 {Sylvain de la Torre$^{1}$\thanks{E-mail: sdlt@roe.ac.uk}
  and Luigi Guzzo$^{2}$ \\
  $^{1}$ SUPA\thanks{Scottish Universities Physics Alliance}, Institute for Astronomy, University of Edinburgh, Royal Observatory, Blackford Hill, EH9 3HJ Edinburgh, UK \\
  $^{2}$ INAF -- Osservatorio Astronomico di Brera, via Emilio Bianchi 46, 23807 Merate, Italy}

\begin{document}

\date{}

\pagerange{\pageref{firstpage}--\pageref{lastpage}} \pubyear{2011}

\maketitle

\label{firstpage}

\begin{abstract}

We investigate the ability of state-of-the-art redshift-space
distortions models for the galaxy anisotropic two-point correlation
function $\xisp$, to recover precise and unbiased estimates of the
linear growth rate of structure $f$, when applied to catalogues of
galaxies characterised by a realistic bias relation. To this aim, we
make use of a set of simulated catalogues at $z=0.1$ and $z=1$ with
different luminosity thresholds, obtained by populating dark-matter
haloes from a large N-body simulation using halo occupation
prescriptions. We examine the most recent developments in
redshift-space distortions modelling, which account for
non-linearities on both small and intermediate scales produced
respectively by randomised motions in virialised structures and
non-linear coupling between the density and velocity fields.  We
consider the possibility of including the linear component of galaxy
bias as a free parameter and directly estimate the growth rate of
structure $f$.  Results are compared to those obtained using the
standard dispersion model, over different ranges of scales.  We find
that the model of \citet{taruya10}, the most sophisticated one
considered in this analysis, provides in general the most unbiased
estimates of the growth rate of structure, with systematic errors
within $\pm 4\%$ over a wide range of galaxy populations spanning
luminosities between $L>L^*$ and $L>3L^*$. The scale-dependence of
galaxy bias plays a role on recovering unbiased estimates of $f$ when
fitting quasi non-linear scales. Its effect is particularly severe for
most luminous galaxies, for which systematic effects in the modelling
might be more difficult to mitigate and have to be further
investigated. Finally, we also test the impact of neglecting the
presence of non-negligible velocity bias with respect to mass in the
galaxy catalogues. This can produce an additional systematic error of
the order of $1-3\%$ depending on the redshift, comparable to the
statistical errors the we aim at achieving with future high-precision
surveys such as Euclid.

\end{abstract}

\begin{keywords}
Cosmology: large-scale structure of Universe -- Galaxies: statistics.
\end{keywords}

\section{Introduction}

The structure in the Universe grows through the competing effects of
universal expansion and gravitational instability. For this reason,
the large-scale spatial distribution and dynamics of galaxies, which
follows in some way those of mass, provide fundamental information
about the expansion history and the nature of gravity. In general,
galaxies are biased tracers of the underlying mass
distribution. However, they are sensitive to the same gravitational
potential and their motions keep an imprint of the rate of structure
growth. One manifestation of this is the observed anisotropy between
the clustering of galaxies along the line-of-sight and that
perpendicular to it in redshift space. These anisotropies or
distortions are caused by the line-of-sight component of galaxy
peculiar velocities affecting the observed galaxy redshifts from which
distances are measured. In turn, the large-scale coherent component of
galaxy peculiar motions is the fingerprint of the growth rate of
structure.

By mapping the large-scale structure over scales which retain this
primordial information, galaxy spectroscopic surveys have become one
of the most powerful probes of the cosmological model. A specific
application of spectroscopic surveys involves recovering cosmological
information on the expansion history $H(z)$, by measuring the shape of
the power spectrum \citep[e.g.][]{tegmark04,cole05,percival07,reid10}
and tracking the baryonic acoustic oscillations (BAO) feature in the
power spectrum or in the two-point correlation function at different
redshifts \citep[e.g][and references
  therein]{eisenstein05,percival10,kazin10,blake11b,anderson12}.
However, measurements of $H(z)$ alone, either from BAO or Type Ia
supernovae, cannot discriminate dark energy from modifications of
General Relativity \citep[e.g.][]{carroll04}, in order to explain the
observed recent acceleration of the expansion of the Universe
\citep{riess98,perlmutter99}. This degeneracy can be lifted by
measuring the growth rate at different epochs
\citep{peacock06,albrecht06}. Indeed, scenarios with similar expansion
history but different gravity or type of dark energy will have a
different rate of structure growth resulting from different effective
gravity strength in action. This makes redshift-space distortions
measured from large spectroscopic surveys a very efficient probe to
test cosmology, at the same level as BAO and cosmological microwave
background (CMB) anisotropies. In fact, although this effect is known
since the late eighties \citep{kaiser87}, its usefulness as a probe of
dark energy and modified gravity has been realised only recently
\citep{guzzo08}.

Measuring the growth rate of structure from redshift-space distortions
is however non trivial. The linear theory formalism for the power
spectrum was first derived by \citet{kaiser87} \citep[see ][for its
  configuration-space counterpart]{hamilton92}. Its validity is
however limited to very large scales as it lacks a description of
small-scale non-linear fluctuations. This model has been extended to
quasi- and non-linear scales in the early nineties using the earlier
ideas of the ``streaming model'' \citep{peebles80}, in which the
linear correlation function is convolved along the line-of-sight with
a pairwise velocity distribution \citep{fisher94,peacock94}. This
enables one to approximately reproduce the Fingers-of-God small-scale
elongation \citep{jackson72}. Fitting functions calibrated on
simulations have also been proposed for this purpose
\citep{hatton99,tinker06}. Such extension of the linear model, usually
referred as the ``dispersion model'', has been extensively used to
measure the growth rate of structure $f$ or the distortion parameter
$\beta=f/b_L$ from redshift surveys, using measurements of both
redshift-space correlation function
\citep{peacock01,hawkins03,zehavi05,ross07,okumura08,guzzo08,daangela08,
  cabre09b,cabre09a,samushia11} and power spectrum
\citep{percival04,tegmark04, tegmark06,blake11}. We refer the reader
to \citet{hamilton98} for a review of older studies.  Although
generally the dispersion model is found to be a good fit on linear and
quasi-linear scales \citep{percival09,blake11}, it breaks down in the
non-linear regime \citep{taruya10,okumura11}. In particular, it has
been shown that it introduces systematic errors of about $10-15\%$ on
the growth rate parameter
\citep[e.g.][]{taruya10,okumura11,bianchi12}, of the order of or
greater than the statistical errors expected from on-going and
prospected very large spectroscopic surveys such as WiggleZ
\citep{drinkwater10}, GAMA \citep{driver11}, VIPERS \citep{guzzo12},
BOSS \citep{white11}, Euclid \citep{laureijs11}, or BigBOSS
\citep{schlegel11}. In particular, Euclid, the recently selected ESA
dark energy mission, should be able to constrain the growth rate at
the percent level \citep{wang10,samushia11b,majerotto12}.  There is
therefore a strong need to go beyond the dispersion model, in order to
bring systematic errors below this expected level of precision,
i.e. measure the growth rate parameter in an \emph{unbiased} way.
This is particularly crucial to be able to disentangle different
models of gravity. For instance, modified-gravity models with Dark
Matter-Dark Energy time-dependent or constant coupling predict
variations from General Relativity on the growth rate smaller than
$10\%$ \citep{guzzo08}.

Work in this direction started since quite some time, concentrating
first on describing the redshift-space clustering and dynamics of dark
matter.  \citet{scoccimarro04} demonstrated that the dispersion model
gives rise to unphysical distributions of pairwise velocities and
proposed an {\sl ansatz} that accounts, to some extent, for the
non-linear coupling between the velocity and the density fields. This
latter model has been shown to provide a better match to the observed
redshift-space power spectrum in dark matter simulations
\citep{scoccimarro04,jennings11} and has later on been refined by
\citet{taruya10}.  Further approaches have also been proposed while
completing this paper \citep{seljak11,reid11,cai12}, but we shall not
discuss them in this analysis.

All mentioned advanced redshift-space distortions models have been
tested so far only on the redshift-space power spectra of dark matter
and dark matter haloes, as extracted from large N-body simulations
\citep{kwan12,reid11,okumura12,nishimichi11}.  This is quite
different from a real survey, in which the most useful tracers of
mass, the galaxies, are in general biased with respect to the
underlying density field through a bias which is generally non-linear
and scale-dependent. The performance of redshift-space distortions
models applied to galaxy populations with \emph{a priori} unknown
biases, has to be precisely investigated. This is the aim of this
paper, in which we confront non-linear models of redshift-space
distortions for the anisotropic two-point correlation function in the
case of realistic galaxy samples. This is done in the framework of the
concordant $\Lambda CDM$ cosmological model. We study the ability of
these models to recover the linear growth rate of structure and their
range of applicability. Furthermore we investigate the effects of
galaxy non-linear spatial and velocity biases and quantify how the
latter affect the estimated linear growth rate for differently biased
galaxy populations.

The paper is organised as follows. In Sect. 2, we present the
redshift-space distortions formalism and how models can be implemented
in practice. In Sect. 3, we present the comparison between the
different models and study the effect of galaxy non-linear bias. In
Sect. 4, we investigate the impact of neglecting galaxy velocity bias
in the modelling. In Sect. 5, we summarise our results and conclude.

\section{Redshift-space distortions theory} \label{sec:rsd}

\subsection{Fourier space}

The peculiar velocity $\bmath{v}$ alters objects apparent comoving
position $\bmath{s}$ from their true comoving position $\bmath{r}$, as
\begin{equation} \label{form:rsdisp}
\bmath{s}=\bmath{r}+\frac{v_\parallel(\bmath{r})\bmath{\hat{e}_\parallel}}{aH(a)},
\end{equation}
where $H(a)$ is the Hubble parameter, $a$ is the scale factor, and
$\bmath{\hat{e}_\parallel}$ is the line-of-sight unit vector. The
redshift-space density field $\delta^s(\bmath{s})$ can be obtained
from the real-space one by requiring mass conservation, i.e.
$\left[1+\delta^s(\bmath{s})\right]d^3\bmath{s}=\left[1+\delta(\bmath{r})\right]d^3\bmath{r}$,
as
\begin{equation} \label{masscons}
\delta^s(\bmath{s})=\left[1+\delta(\bmath{r})\right]\left|\frac{d^3\bmath{s}}{d^3\bmath{r}}\right|^{-1}-1.
\end{equation}
In the following we shall work in the plane-parallel approximation and
in this limit, the Jocobian of the real- to redshift-space
transformation can be simply written as,
\begin{equation} \label{jacobian}
\left|\frac{d^3\bmath{s}}{d^3\bmath{r}}\right|=1-f\partial_\parallel u_\parallel,
\end{equation}
where we defined $u_\parallel(\bmath{r})=-v_\parallel(\bmath{r})/(f
aH(a))$ with $f$ being the linear growth rate. The linear growth rate
parameter is defined as the logarithmic derivative of the linear
growth factor $D(a)$ and given by $f(a)=d\ln{D}/d\ln{a}$. To a very
good approximation it has a generic form \citep{wang98,linder05},
\begin{equation}
\label{eq:fgamma} f(a)\simeq\Omega_m(a)^\gamma 
\end{equation}
where
\begin{equation}\label{eq:omegaa}
\Omega_m(a)=\frac{\Omega_{m,0}}{a^3}\frac{H_0^2}{H^2(a)} .
\end{equation}
In this parametrisation, while $\Omega_m$ characterises the mass
content in the Universe, the exponent $\gamma$ directly relates to the
theory of gravity \citep[e.g.][]{linder04}. General Relativity
scenarios have $\gamma\simeq0.55$.

From Eq. \ref{masscons} and Eq. \ref{jacobian}, one can write the
redshift-space density field as,
\begin{equation}
\delta^s(\bmath{s})=\left(\delta(\bmath{r})+f\partial_\parallel
u_\parallel\right)\left(1-f\partial_\parallel u_\parallel\right)^{-1}. \label{rsdelt0}
\end{equation}
One usually assumes an irrotational velocity field for which
$u_\parallel(\bmath{r})=\partial_\parallel
\Delta^{-1}\theta(\bmath{r})$ and where
$\theta(\bmath{r})=\bmath{\nabla} \cdot \bmath{v}(\bmath{r})$ is the
velocity divergence field and $\Delta$ denotes the Laplacian. In that
case Eq. \ref{rsdelt0} can be recast,
\begin{equation}
\delta^s(\bmath{s})=\left(\delta(\bmath{r})+f\partial_\parallel^2\Delta^{-1}\theta(\bmath{r})\right)\left(1-f\partial_\parallel^2\Delta^{-1}\theta(\bmath{r})\right)^{-1}.
\end{equation}
In Fourier space, it is noticeable that $\partial_\parallel^2
\Delta^{-1}=(k_\parallel/k)^2=\mu^2$ with $\mu$ being the cosine of the angle
between the line-of-sight and the separation vector. Therefore, one
can write the redshift-space density field \citep{scoccimarro99} as,
\begin{align}
\delta^s(k,\mu) &={} \int \frac{d^3\bmath{s}}{(2\pi)^3} e^{-i\bmath{k} \cdot \bmath{s}}\delta^s(\bmath{s}) \\
&={} \int \frac{d^3\bmath{r}}{(2\pi)^3} e^{-i\bmath{k} \cdot \bmath{r}}e^{-ikf\mu} [\delta(\bmath{x})+\mu^2 f \theta(\bmath{x})] \label{rsdelt}
\end{align}
and the redshift-space power spectrum as,
\begin{align}
P^s(k,\mu)&={}\int \frac{d^3\bmath{r}}{(2\pi)^3} e^{-i\bmath{k} \cdot \bmath{r}}\left<e^{-ikf\mu \Delta u_\parallel} \times \right. \nonumber \\
& \left. [\delta(\bmath{x})+\mu^2 f \theta(\bmath{x})][\delta(\bmath{x}^\prime)+\mu^2 f \theta(\bmath{x}^\prime)]\right> \label{rspk}
\end{align}
where in the latter equation, $\Delta
u_\parallel=u_\parallel(\bmath{x})-u_\parallel(\bmath{x}^\prime)$ and
$\bmath{r}=\bmath{x}-\bmath{x}^\prime$. The redshift-space power
spectrum given in Eq. \ref{rspk} is almost exact, the only
approximation which has been done is to assume that all object
line-of-sight separations are parallel. This approximation is valid
for samples with pairs covering angles typically lower than $10^\circ$
\citep{matsubara00}. Eq. \ref{rspk} captures all the different regimes
of distortions. While the terms in square brackets describe the
squashing effect or ``Kaiser effect'' which leads to an enhancement of
clustering on large scales due to the coherent infall of mass towards
overdensities, the exponential prefactor is responsible to some extent
for the Fingers-of-God effect \citep[FoG,][]{jackson72} which
disperses objects along the line-of-sight due to random motions in
virialised structures.  \citet{scoccimarro04} proposed a simple ansatz
for the redshift-space anisotropic power spectrum by making the
assumption that the exponential prefactor and the term involving the
density and velocity fields can be separated in the ensemble
average. In that case Eq.  \ref{rspk} simplifies to,
\begin{equation}
P^s(k,\mu)= e^{-(fk\mu\sigma_v)^2}\left[\Pdd(k)+2\mu^2 f
\Pdt(k)+\mu^4 f^2 \Ptt(k)\right], \label{rspksco}
\end{equation}
where $\Pdd$, $\Pdt$, $\Ptt$ are
respectively the non-linear mass density-density, density-velocity
divergence, and velocity divergence-velocity divergence power spectra
and $\sigma_v$ is the pairwise velocity dispersion defined as,
\begin{equation}
\sigma^2_v=\frac{1}{6\pi^2}\int \Ptt(k)dk. \label{sigmav}
\end{equation}
It is found that this model captures most of the distortion features
predicted by N-body simulations \citep{scoccimarro04,jennings11}
although it breaks down in the non-linear regime
\citep{percival09,taruya10}. Note that in the linear regime where
$\Pdd=\Pdt=\Ptt=P$ and in the limit where $k\sigma_v$ tends to zero,
one recovers the original \citet{kaiser87} formula,
\begin{equation} \label{kaisermodel}
P^s(k,\mu) = [1 + 2\mu^2f + \mu^4f^2]P(k),
\end{equation}
derived from linear-order calculations.

In principle, the exponential prefactor and the term involving the
density and velocity fields in Eq. \ref{rspk}, which we will refer to
as the damping and Kaiser terms in the following, cannot be treated
separately. Additional terms may arise in Eq. \ref{rspksco} from the
coupling between the exponential prefactor and the velocity divergence
and density fields. \citet{taruya10} proposed an improved model that
takes into account these couplings, adding two correction terms $C_A$
and $C_B$ to \citet{scoccimarro04}'s formula such as,
\begin{align}
P^s(k,\mu) &={} D(k\mu\sigma_v)\left[\Pdd(k)+2\mu^2 f \Pdt(k)+\mu^4 f^2 \Ptt(k) \right. \nonumber \\
&\left. + C_A(k,\mu;f)+C_B(k,\mu;f)\right], \label{rspktar}
\end{align}
whose perturbative expressions are given in their appendix A. In the
improved model, the exponential prefactor has been replaced by an
arbitrary functional form $D(k\mu\sigma_v)$ for which $\sigma_v$ is an
\emph{effective} pairwise velocity dispersion parameter that can be
fitted for. \citet{taruya10} showed that while adopting a Gaussian or
a Lorentzian for the damping function and letting $\sigma_v$ free, one
improves dramatically the fit to the redshift-space power spectrum in
large dark matter simulations, particularly on translinear scales.

The function $D(k\mu\sigma_v)$ damps the power spectra in the Kaiser
term but also partially mimics the effects of the pairwise velocity
distribution (PVD) in virialised systems, which translate into the FoG
seen in the anisotropic power spectrum and correlation function on
small scales. This is analogous to the phenomenological dispersion
model proposed in the early nineties
\citep[e.g.][]{fisher94,peacock94} in which the linear Kaiser model in
configuration space \citep{hamilton92} is radially convolved with a
PVD model to reproduce the FoG elongation on small scales, as for the
early streaming model \citep{peebles80}.

There is however not any simple general functional form for the PVD
that matches all scales for all types of tracers. The shape of the PVD
is found to depend on galaxy physical properties and halo occupation
\citep{li06,tinker06}, and its associated pairwise velocity dispersion
to vary with scale, in particular at small separations
\citep[e.g.][]{hawkins03,cabre09a}. It can be shown mathematically
that the PVD is in fact not a single function but rather an infinite
number of PVD corresponding to different scales and angles between
velocities and separation vectors \citep{scoccimarro04}. In practice
however, the use of an exponential distribution, a Gaussian or other
forms with more degrees of freedom \citep[e.g.][]{tang11,kwan12} shows
to be very useful to fit the residual small-scale distortions
remaining once the large-scale Kaiser distortions are accounted for,
unless one is interested in modelling the very small highly non-linear
scales.

\subsection{Configuration space}

The redshift-space anisotropic two-point correlation function can be
obtained by Fourier-transforming the anisotropic redshift-space power
spectrum as,
\begin{equation} 
\xisp = \int \frac{d^3\bmath{k}}{(2\pi)^3} e^{i\bmath{k}\cdot\bmath{s}}
P^s(k,\mu) = \sum_{l} \xi^s_l(s)L_l(\nu)
\end{equation}
where $\nu=\rpa/s$, $\rpe=\sqrt{s^2-\rpa^2}$, and $L_l$ denote Legendre
polynomials. The correlation function multipole moments
$\xi^s_l(s)$ are defined as,
\begin{equation} \label{expmom}
\xi^s_l(s)=i^l \int \frac{dk}{2\pi^2} k^2 P^s_l(k)j_l(ks),
\end{equation} 
where $j_l$ denotes the spherical Bessel functions and
\begin{equation} \label{expmomK}
P^s_l(k)=\frac{2l+1}{2} \int_{-1}^1 d\mu P^s(k,\mu) L_l(\mu).
\end{equation}
In practice, it can be convenient to write the redshift-space
two-point correlation function as a convolution between the Fourier
transform of the damping function $D$ and that of the Kaiser term as,
\begin{align}
\xisp &={} \hat{D}(\rpa,\sigma_{v}) \otimes \int \frac{d^3\bmath{k}}{(2\pi)^3} e^{i\bmath{k}\cdot\bmath{s}} P_K^{s}(k,\mu) \nonumber \\
&={} \hat{D}(\rpa,\sigma_{v}) \otimes \xi_K\left(\rpe, \rpa\right) \nonumber \\
&={} \hat{D}(\rpa,\sigma_{v}) \otimes \sum_l \xi^{s,K}_l(s)L_l(\nu)
\end{align}
In the case of the model of Eq. \ref{rspksco}, the three non-null
correlation function multipole moments related to the Kaiser term are
given by,
\begin{align} 
\xi^{s,K}_0(s) &={} \xidd(r) + f\frac{2}{3}\xidt(r) + f^2\frac{1}{5}\xitt(r) , \\ 
\xi^{s,K}_2(s) &={} f\frac{4}{3}\xi^{(2)}_{\delta\theta}(r) + f^2\frac{4}{7}\xi^{(2)}_{\theta\theta}(r) , \\
\xi^{s,K}_4(s) &={} f^2\frac{8}{35}\xi^{(4)}_{\theta\theta}(r),
\end{align}
where $\xidd$, $\xidt$, $\xitt$ are the Fourier conjugate pairs of
$\Pdd$, $\Pdt$, $\Ptt$ and \citep{hamilton92,cole94},
\begin{align}
\xi^{(2)}_{X}(r)&={} \xi_{X}(r) - \frac{3}{r^3} \int_0^r\xi_{X}(r')r'^2dr' \\
\xi^{(4)}_{X}(r)&={} \xi_{X}(r) + \frac{5}{2} \frac{3}{r^3} \int_0^r\xi_{X}(r')r'^2dr' \nonumber \\
& - \frac{7}{2} \frac{5}{r^5} \int_0^r\xi_{X}(r')r'^4dr' .
\end{align}
The correlation function multipole moments of the Kaiser term in the
case of the model of Eq. \ref{rspktar} are given in Appendix A. For
the latter model we restrain ourselves to use only the first three
non-null multipole moments, as those of orders $l=6$ and $l=8$ are
very poorly defined in our simulated galaxy catalogues.

\subsection{From mass to galaxies}

The models derived in the previous section apply in the case of
perfectly unbiased tracers of mass. Real galaxies however are biased
with respect to mass. Galaxy biasing is generally expected to be
non-linear, scale-dependent, stochastic, and to depend on galaxy type,
although it is still poorly constrained by observations. On large
scales in the linear regime, one expects the bias to be a constant
multiplicative factor to the mass density field as
$\delta_g=b_L\delta$. In that case, it is convenient to replace the
growth rate $f$ in the models with a ``effective'' growth rate (or
distortion parameter) $\beta=f/b_L$, which accounts for the
large-scale linear bias $b_L$ of the considered galaxies. This simple
model is valid on large scales where the bias asymptotes to a constant
value but breaks down on small non-linear scales, where bias possibly
varies with scale. Recently, \citet{okumura11} showed that the
scale-dependent behaviour of halo bias can strongly affect the
recovery of the growth rate. While some analytical approaches have
been proposed to include bias non-linearity in the model
\citep{desjacques10,matsubara11}, here we follow a different route and
assume that the scale dependence of bias is known. In fact, the latter
can be measured to some extent from the data themselves in
configuration space, once the shape for the underlying non-linear mass
power spectrum is assumed.  General arguments may suggest that galaxy
motions are also biased with respect to the mass velocity field, while
observations tend to indicate that this bias is small
\citep{tinker06,skibba11}. In this analysis we will neglect the galaxy
velocity bias in the models but discuss and quantify its impact on the
recovery of $f$ in Section \ref{secgalvelbias}.

\subsection{Constructing the galaxy redshift-space distortion models} \label{sec:models}

We will use in this analysis different combinations of Kaiser terms,
damping functions, and bias prescriptions. Although we will work in
configuration space, we refer to the different models in this section
as their Fourier-space counterpart for clarity. All the models we
consider take the general form,
\begin{equation}
P_g^s(k,\mu)=D(k\mu\sigma_v)P_K(k,\mu,b)
\end{equation}
where,
\begin{align}
D(k\mu\sigma_v)=\left\{ \nonumber
\begin{array}{lcl}
\exp(-(k\mu\sigma_v)^2)
\\
\\
1/(1+(k\mu\sigma_v)^2)
\end{array}
\right.
\end{align}
and,
\begin{align}
P_K(k,\mu,b) = \hspace{4cm} & \nonumber \\ 
\left\{ \nonumber
\begin{array}{lcl}
b^2(k) \Pdd(k)+2\mu^2 fb(k) \Pdd(k) +\mu^4 f^2 \Pdd(k) \;\;\;\;  {\rm (mod.~A)} \\
\\
b^2(k) \Pdd(k)+2\mu^2 fb(k) \Pdt(k) +\mu^4 f^2 \Ptt(k) \;\;\;\; {\rm (mod.~B)}  \\
\\
b^2(k) \Pdd(k)+2\mu^2 fb(k) \Pdt(k) +\mu^4 f^2 \Ptt(k) \\ 
+ C_A(k,\mu;f,b) + C_B(k,\mu;f,b) \;\;\;\; \;\;\;\; \;\;\;\; \;\;\;\; \;\;\;\; \;\;\;\; \; {\rm (mod.~C)} 
\end{array}
\right.
\end{align}

\begin{align}
b(k)=\left\{ \nonumber
\begin{array}{lcl}
b_L
\\
\\
b_L b_{NL}(k)
\end{array}
\right. \;\; .
\end{align}

Hereafter, we will refer as the different $P_K$ models to A, B, and
C. Model A corresponds to the \citet{kaiser87} model with the
non-linear power spectrum instead of the linear one. It assumes a
linear coupling between the density and velocity fields such that
$\delta\propto\theta$. Model B is the generalisation proposed by
\citet{scoccimarro04} that accounts for the non-linear coupling
between the density and velocity fields, making explicitly appearing
the velocity divergence auto-power spectrum and density--velocity
divergence cross-power spectrum. Finally, model C is an extension of
model B that contains the two additional correction terms proposed by
\citet{taruya10} to correctly account for the coupling between the
Kaiser and damping terms. Besides, we will consider two deterministic
galaxy biasing prescriptions: a constant linear bias $b(k)=b_L$ and a
general non-linear bias which we define as
$b(k)=\left(P_{gg}/\Pdd\right)^{1/2}(k)=b_Lb_{NL}(k)$, where $P_{gg}$
is the galaxy power spectrum and $b_{NL}(k)$ is the scale-dependent
part of the bias that tends to unity at small $k$. We note that the
Gaussian and Lorentzian damping forms that we will consider here,
correspond respectively to Gaussian and exponential functions in
configuration-space.

The redshift-space distortions models necessitate $\Pdd$, $\Pdt$, and
$\Ptt$ real-space power spectra as input. These can be obtained
analytically using perturbation theory. Although standard perturbation
theory does not describe well the shape of these power spectra on
intermediate and non-linear scales, improved treatments such as
Renormalised Perturbation Theory \citep[RPT,][]{crocce06} or Closure
Theory \citep{taruya09} have been shown to be much more accurate
\citep[see][for a thorough comparison]{carlson09}. In particular,
Closure Theory predictions are found to match large N-body simulation
real-space power spectra to the percent-level up to $k=0.2$ for
$z>0.5$ \citep{taruya09}.  

\begin{figure}
\includegraphics[width=84mm]{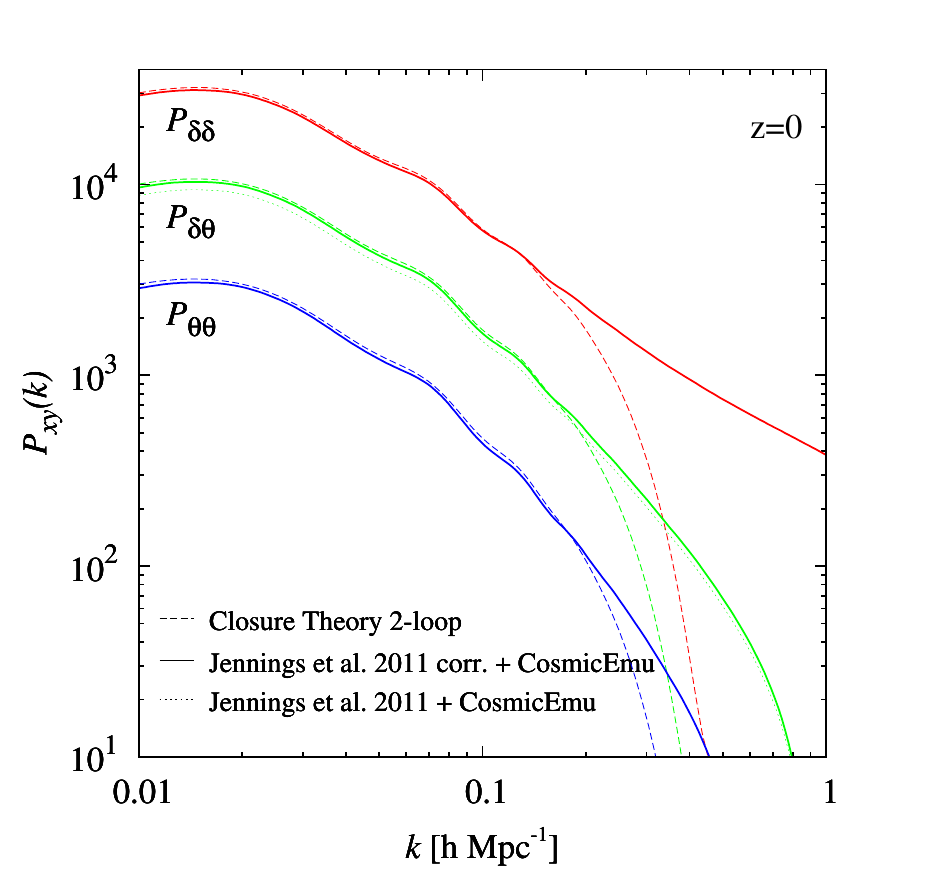}
\caption{The predicted $\Pdd$, $\Pdt$, $\Ptt$ by CosmicEmu
  \citep{lawrence10} and \citet{jennings11} fitting functions (solid
  and dotted curves), and Closure theory \citep{taruya09} (dashed
  curves) at $z=0$ in the assumed cosmology. In all cases, $\Pdt$ and
  $\Ptt$ have been divided by a factor of 3 and 10 respectively to
  improve the clarity of the figure.}
\label{comppk0}
\end{figure}

\begin{figure}
\includegraphics[width=84mm]{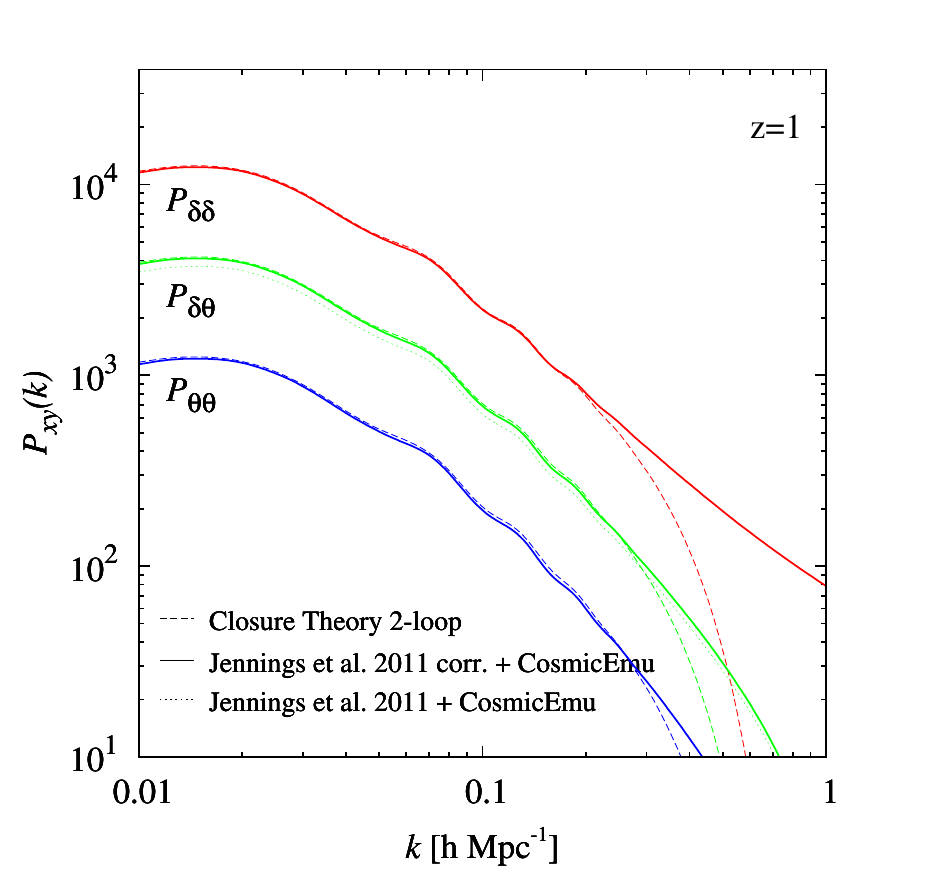}
\caption{Same as Fig. \ref{comppk0} but for $z=1$.}
\label{comppk1}
\end{figure}

In this analysis we use the $\Pdd$ provided by CosmicEmu emulator
\citep{lawrence10} and the fitting functions of \citet{jennings11} to
obtain $\Ptt$ and $\Pdt$ from $\Pdd$. The latter fitting functions
have an accuracy of $5\%$ to $k=0.2$ for both standard $\Lambda CDM$
and quintessence dark energy cosmological models. In
Fig. \ref{comppk0} and \ref{comppk1} we compare the $\Pdd$, $\Pdt$,
$\Ptt$ obtained in this way with Closure Theory 2-loop analytical
predictions at $z=0$ and $z=1$. We find that all power spectra agree
very well below $k\simeq0.2$ and $k\simeq0.3$ respectively for the two
redshifts considered, except in the case of $\Pdt$ for which they
systematically differ by about $10\%$. While all other power spectra
match on linear scales, the $\Pdt$ fitting formula from
\citet{jennings11} stays somewhat below (dotted lines in the
figures). We find that by multiplying the latter by a factor of $1.1$
one obtains a good match with Closure Theory predictions on both
linear and non-linear scales (solid lines in the figures). We will
then adopt this correcting factor in the following when calculating
the redshift-space distortions models.

It is noticeable that Closure Theory breaks down at lower $k$ than
\citet{jennings11} fitting functions, indicating that the latter are
more suitable in practice to describe clustering on the smallest
scales. It is however important to mention that the validity of these
fitting functions is limited at large $k$
\citep[$k_{max}=0.2-0.3$,][]{jennings11}.  One can therefore reliably
describe the models involving $\Pdt$ and $\Ptt$ only down to scales of
$r\simeq \pi/k_{max}\simeq 10.5~\mhmpc$. On smaller scales, $\Pdt$ and
$\Ptt$ configuration-space counterparts, $\xidt$ and $\xitt$, will
drop rapidly and their contributions to the redshift-space distortions
models will be underestimated with respect to that of $\xidd$.

\section{Model testing}

\subsection{Methodology}

To test the redshift-space distortions models presented in section
\ref{sec:models} and quantify how well the linear growth rate
parameter $f$ can be recovered from the anisotropic two-point
correlation function $\xisp$, we constructed a set of realistic galaxy
catalogues. We populate the identified friends-of-friends haloes in
the MultiDark Run 1 (MDR1) dark matter N-body simulation
\citep{prada11} with galaxies by specifying the Halo Occupation
Distribution (HOD). MDR1 assumes a $\Lambda CDM$ cosmology with
$(\Omega_m=\Omega_{dm}+\Omega_b,~\Omega_\Lambda,~\Omega_b,~h,~n,~\sigma_8)
= (0.27,~0.73,~0.0469, ~0.7,~0.95,~0.82)$ and probe a cubic volume of
$1~{\rm h}^{-3}~{\rm Gpc}^3$ with a mass resolution of
$m_p=8.721\times10^9~\rm{h^{-1}}~\rm{M_\odot}$. From haloes at
snapshots $z=0.1$ and $z=1$, we built galaxy catalogues based on the
current most accurate observations of the halo occupation (HOD)
available at these redshifts \citep{zheng07,zehavi11,coupon12} and
create three luminosity-threshold samples corresponding to $L>L^*$,
$L>2L^*$, and $L>3L^*$. In these catalogues, the redshift-space
displacements with respect to real space were reproduced using
Eq. \ref{form:rsdisp} and the galaxy peculiar velocity
information. Consistently with model assumptions, we used the
plane-parallel approximation and applied redshift-space distortions
along one dimension of the simulation boxes only. The details of the
procedure used to create the galaxy catalogues are given in Appendix
B. The main limitation of using HOD for this study rely on the
hypothesis of halo sphericity and isotropy. These assumptions can have
an influence on the dynamics of galaxies inside haloes, in particular
on their random velocities. However, these have only a very limited
impact on this analysis which focuses on scales greater than $1$\hmpc,
where the effect of possible anisotropies in the galaxy velocity
distribution in haloes is only marginal.

We measured the anisotropic two-point auto-correlation function
$\xisp$ in the redshift-space catalogues using the \citet{landy93}
estimator in linearly-spaced bins of $0.5\mhmpc$ in both $\rpe$ and
$\rpa$ directions. Because of the large number of galaxies and to keep
the computational time reasonable, all pair counts have been performed
using a specifically developed parallel \emph{kd-tree} code following
the \emph{dual-tree} approach \citep{moore01}.

The best-fitting parameters for the different models have been
determined by adopting the usual likelihood function,
\begin{equation}
  -2\ln{\mathcal{L}}=\sum_{i=1}^{N_p}\sum_{j=1}^{N_p}\Delta_i C^{-1}_{ij} \Delta_j
\end{equation}
where $N_p$ is the number of points in the fit, $\Delta$ is the
data-model difference vector, and $C$ is the covariance matrix of the
data. The likelihood is performed on the quantity
$y=\ln\left(1+\xisp\right)$, rather than simply $\xisp$, as to enhance
the weight on large more linear scales \citep[see][for
  discussion]{bianchi12}.

\begin{figure}
\includegraphics[width=84mm]{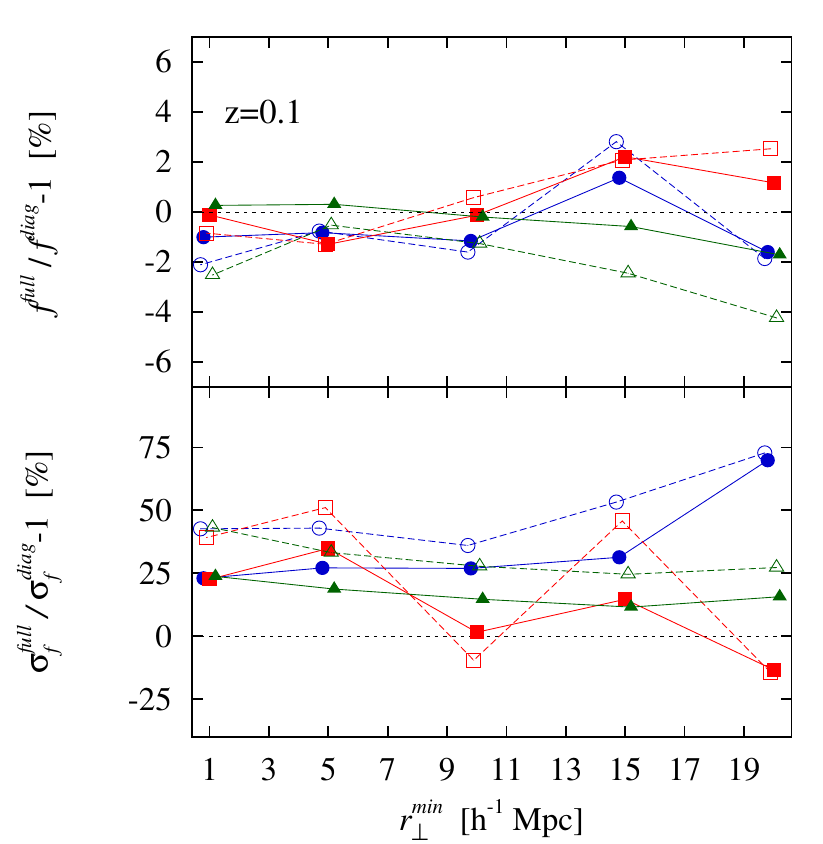}
\caption{Relative systematic (top) and statistical (bottom) errors on
  the growth rate parameter for $L>L^*$ galaxies at $z=0.1$ when the
  full covariance matrix based on 64 jackknife resamplings is used in
  the fitting or only its diagonal elements. The different symbols
  correspond to the different models quoted in Fig. \ref{systz1},
  except that empty symbols here corresponds to models with
  scale-dependent bias while filled corresponds to those with linear
  bias. In all cases we assumed an exponential form for the damping
  term in the models.}
\label{syst_com}
\end{figure}

The determination of the covariance matrix is however troublesome when
fitting two-dimensional correlation functions. Having only one
realisation of the simulated catalogues at each redshift, we can use
to this end internal estimators, such as blockwise bootstrap or
jackknife resampling \citep[e.g.][]{norberg09}. Using the latter
method, we find that the maximum number of cubic sub-volumes that can
be extracted without underestimating the variances on the scales of
interest is 64. This poses a problem, since in order to have a proper
estimation of the eigenvalues of the covariance matrix, this number
should be at least equal to the number of degrees of freedom, which in
our case ranges between 14397 and 25597 (i.e. $120^2$ to $160^2$ data
points minus $3$ free parameters) depending on the scale interval of
the fit.  As a result, the covariance matrix estimated in this way is
degenerate and in the end not very useful. We note that this is a
general problem for redshift-space distortions analysis, when one
tries to fit the full anisotropic two-point correlation function or
power spectrum. In principle this could be solved by using a very
large number of realisations or alternatively, theoretically-motivated
analytical forms for the covariance matrix. In our case, due to the
limited size of the simulation, it is impracticable to define more
sub-volumes than degrees of freedom, as this would result in
underestimating the covariances by using too small sub-volumes. We
note however that by estimating $\xisp$ using bins of size $0.5\mhmpc$
in both directions in the jackknife resamplings, we are oversampling
the functions so the actual number of degrees of freedom may be
smaller than that associated to the number of data points
\citep[e.g.][]{fisher94}. We are therefore forced to perform our fits
ignoring the non-diagonal elements of the covariance, i.e. use the
variances only. We verified however on a test case that the best-fit
values of the parameters obtained in this way do not differ
significantly when using the full covariance matrix based on 64
sub-volumes. This is presented in Fig. \ref{syst_com} where it is
shown that for $L>L^*$ galaxies at $z=0.1$, the recovered value of $f$
do not differ by more than $1-2\%$. We note that statistical errors
may however be underestimated by up to about $50\%$ when not using the
full covariances.

In all cases we define $f$, $\sigma_v$, and $b_L$ as free parameters
and use different scale ranges in the fit by varying $\rpe$ from
$\rpe^{min}=1\mhmpc$ to $\rpe^{min}=20\mhmpc$ and fixing
$\rpe^{max}=\rpa^{max}=80\mhmpc$. The statistical errors on the model
parameters, and in particular on $f$, have been estimated from the
$1\sigma$ dispersion of their best-fitted values among the $64$
resamplings. Because our main aim is to compare the accuracy of
different models of redshift-space distortions, we assume the shape
and normalisation ($\sigma_8$) of the input mass power spectra to be
perfectly known and fix them to those of the simulation. We use for
the growth rate fiducial values $f^{fid}$ those given by
Eq. \ref{eq:fgamma}.

\subsection{Varying the Kaiser and damping terms}

\begin{figure}
\includegraphics[width=84mm]{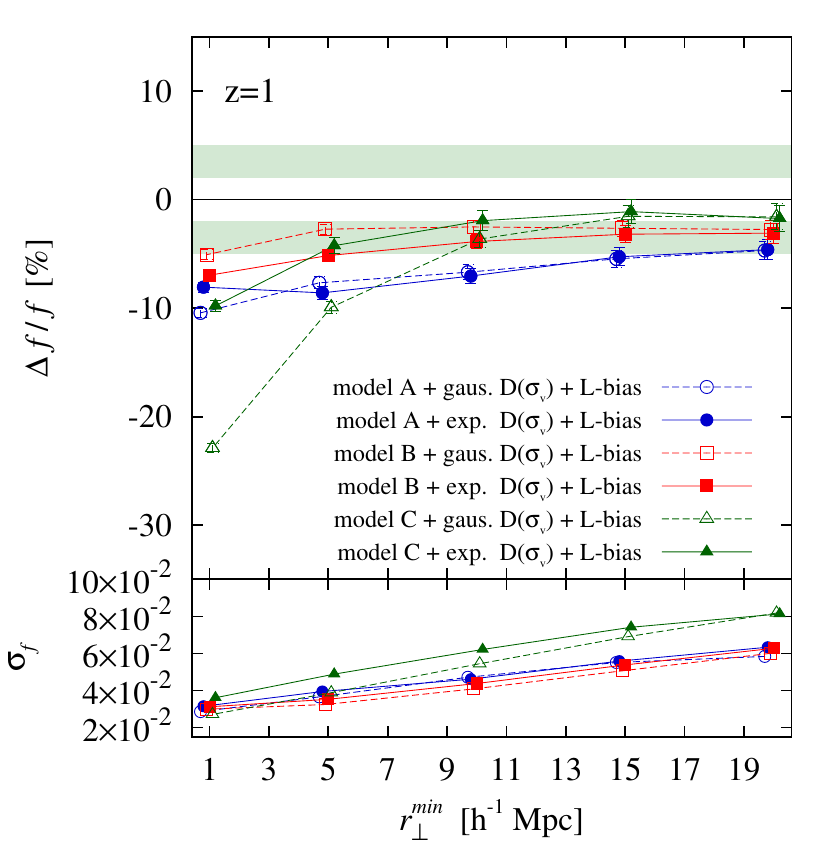}
\caption{Top: relative systematic error on $f$ and its
  corresponding 1$\sigma$ uncertainty for $L>L^*$ galaxies at $z=1$,
  in the case of models for which galaxy bias is assumed to be
  linear. The light (dark) shaded band marks the $2\%$ ($5\%$) region
  around the fiducial value. Bottom: 1$\sigma$ statistical error on
  $f$ in the case of a survey probing $1~{\rm h}^{-3}~{\rm Gpc}^3$.}
\label{systz1}
\end{figure}

\begin{figure}
\includegraphics[width=84mm]{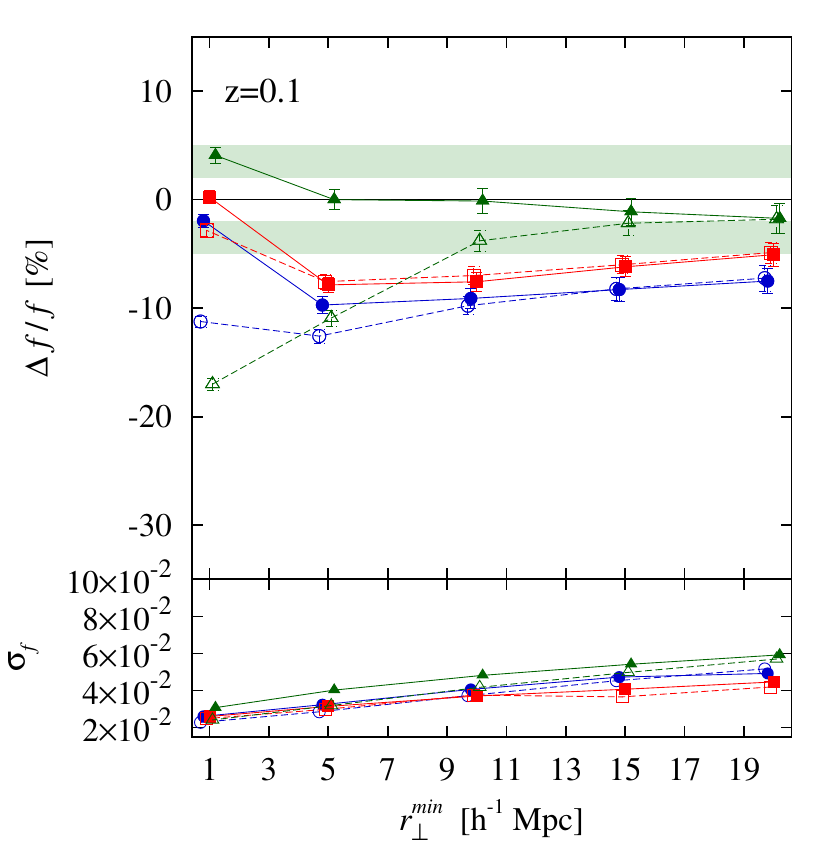}
\caption{Same as Fig. \ref{systz1} but at redshift $z=0.1$.}
\label{systz0}
\end{figure}

Let us first study the effect of using different combinations of
Kaiser term and damping function with the assumption that galaxy bias
is linear and scale-independent. We use the full galaxy catalogues at
$z=1$ and $z=0.1$ and estimate the statistical and systematic errors
on the growth rate with the different models, varying the minimum
perpendicular scale used in the fit, $\rpe$. In this and the following
section, we consider simulated galaxies with luminosities $L>L^*$ (see
Appendix B for details), having linear biases of $b_L=1.34$ and
$b_L=2.01$ respectively at $z=0.1$ and $z=1$. These values are
realistically close to current observations
\citep[e.g.][]{norberg01,pollo05,coil06,zehavi11}.

The different models behave quite similarly at both $z=1$ and $z=0.1$,
as shown respectively in Fig.~\ref{systz1} and Fig.~\ref{systz0}. The
systematic errors on the growth rate are significant using any of the
model variants, and depends on the chosen minimum scale for the fit,
$\rpe^{min}$. Although in general all models tend to underestimate the
growth rate, the systematic errors gradually diminishes while passing
from model A to model C, the latter performing best. In particular, in
the case of model C with exponential damping (\cexp), $|\Delta f/f|$
always remains below $2\%$ at both redshift $z=1$ and $z=0.1$ for
$\rpe^{min}>10\mhmpc$. Model B perform substantially worse,
underestimating the growth rate by $3-7\%$ and $5-8\%$ at $z=1$ and
$z=0.1$ respectively. Finally, we note that model A with exponential
damping (\aexp) applied to scales $\rpe^{min}<10\mhmpc$, which is one
of the most commonly used models in the literature, performs worst,
systematically underestimating $f$ by up to $10\%$ in agreement with
recent analysis \citep[e.g.][]{bianchi12}.

These results are qualitatively consistent with the power spectrum
analysis of \citet{kwan12}, who show that for dark matter only at
$z=0$, $z=0.5$ and $z=1$, model C with Gaussian damping \footnote{This
  model is referred to as \emph{Taruya++ with empirical damping} in
  \citet{kwan12}} (\cgauss) is the least biased model when fitting up
to $k_{max}=0.1$. Our tests show however that for galaxies, model
\cexp is less biased than \cgauss. In fact, the choice of damping
function has only a significant impact on the model's ability to
handle small scales, with the difference diminishing with increasing
$\rpe^{min}$ given the similar asymptotic behaviour of the two
functional forms. Conversely, we note that the Gaussian damping
produces in general slightly lower statistical errors than the
exponential damping. These tend also to be about $15\%$ smaller for
models A and B than for model C.

It is important to note that for $\rpe^{min}<10\mhmpc$, the accuracy
with which $f$ is recovered tends to deteriorate for all models. This
is plausibly associated with the increase of non-linearities in galaxy
clustering. In this regime, the assumption of linear biasing breaks
down and it becomes crucial to account for non-linearities to recover
unbiased measurements of the growth rate, as we will discuss in the
next sections.

\begin{figure}
\includegraphics[width=80mm]{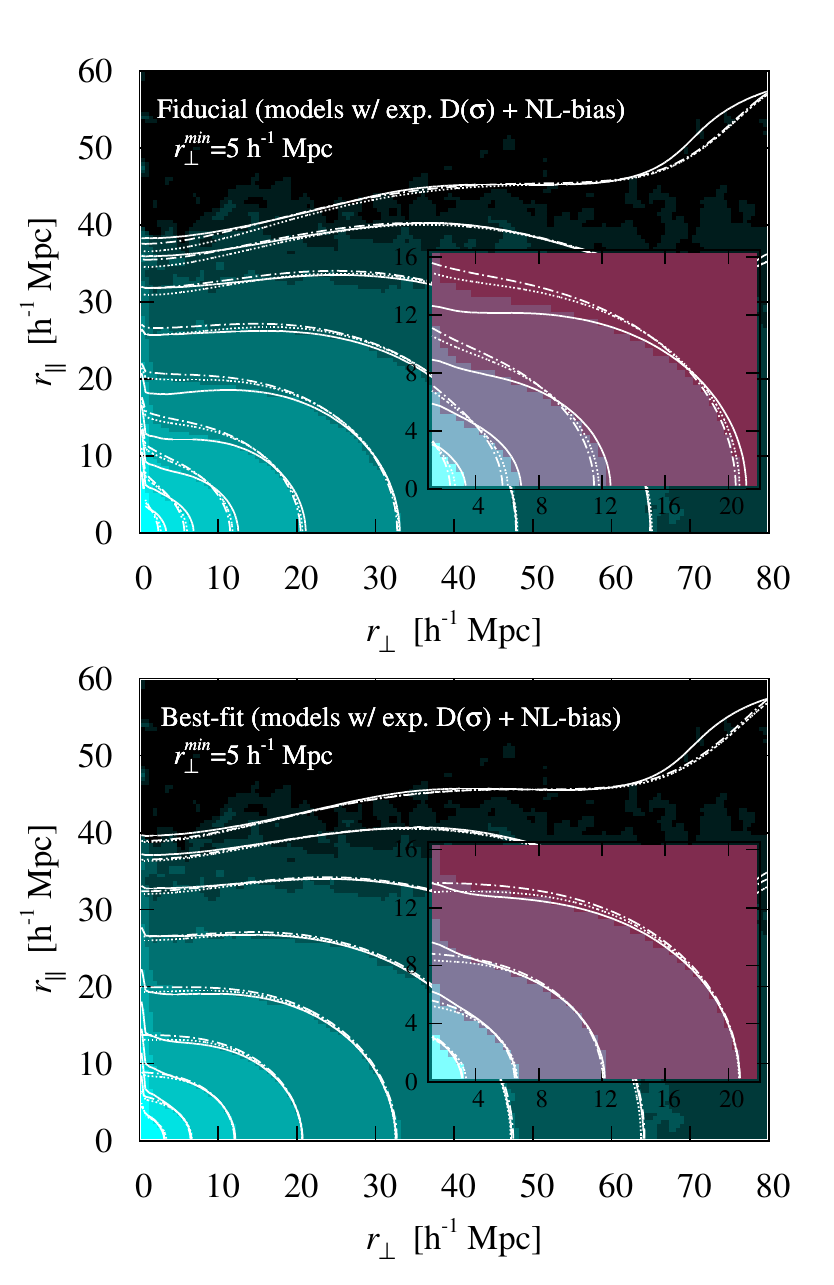}
\caption{Measured $\xisp$ and associated models for $L>L^*$ galaxies
  at $z=1$. In each panel the dotted, dot-dashed, and solid curves
  correspond respectively to model A, B, and C with exponential
  damping and linear bias, while the contours correspond to the
  measured $\xisp$ in the galaxy catalogue. The top panel shows the
  fiducial prediction of the models while the bottom panel shows the
  best-fitting model when the parameters ($f$,$\sigma_v$,$b_L$) are
  allowed to vary. We note the fiducial value for $\sigma_v$ is fixed
  to its linear value. In this figure, the measured $\xisp$ is
  smoothed using a Gaussian kernel of size $0.5\mhmpc$.}
\label{contours_z1}
\end{figure}

\begin{figure}
\includegraphics[width=80mm]{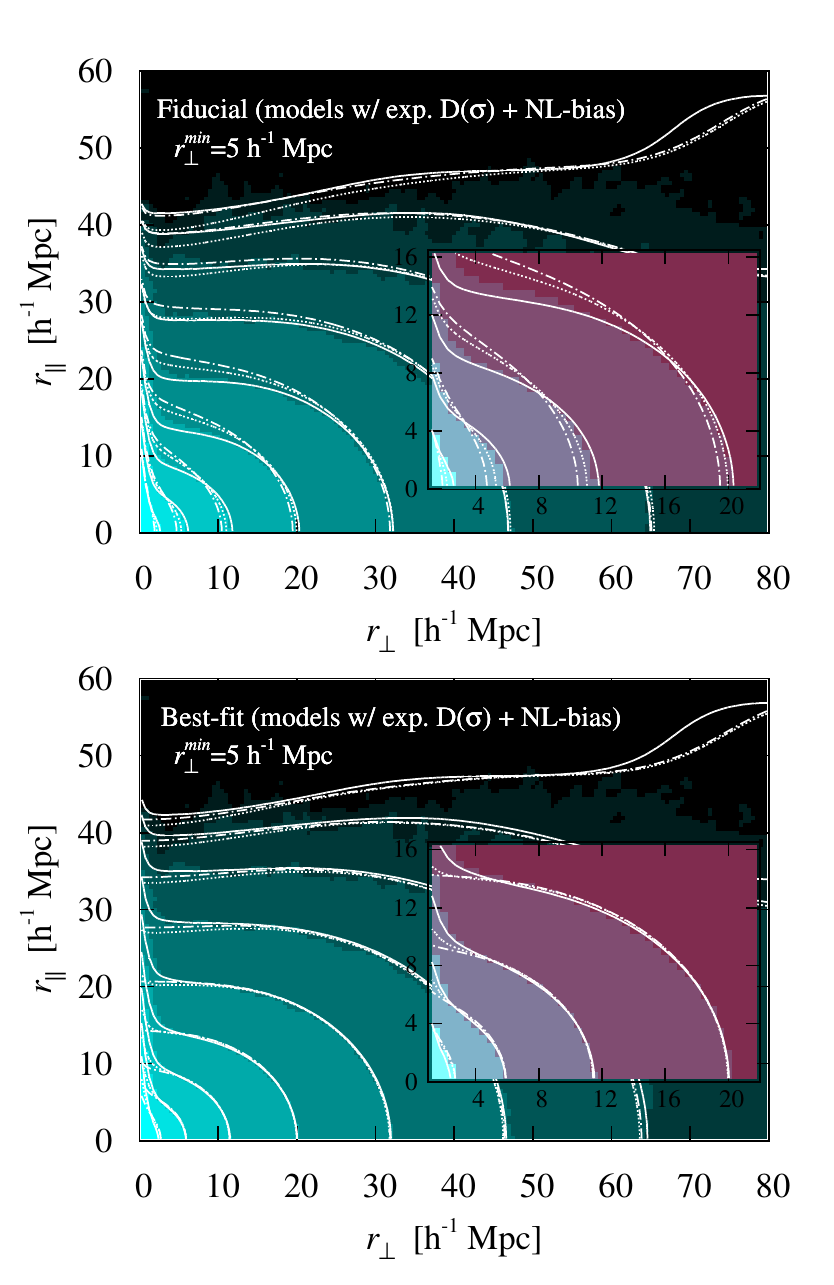}
\caption{Same as Fig. \ref{contours_z1} but at $z=0.1$.}
\label{contours_z0}
\end{figure}

\subsection{Effect of galaxy scale-dependent bias} \label{secgalnlbias}

\begin{figure}
\includegraphics[width=84mm]{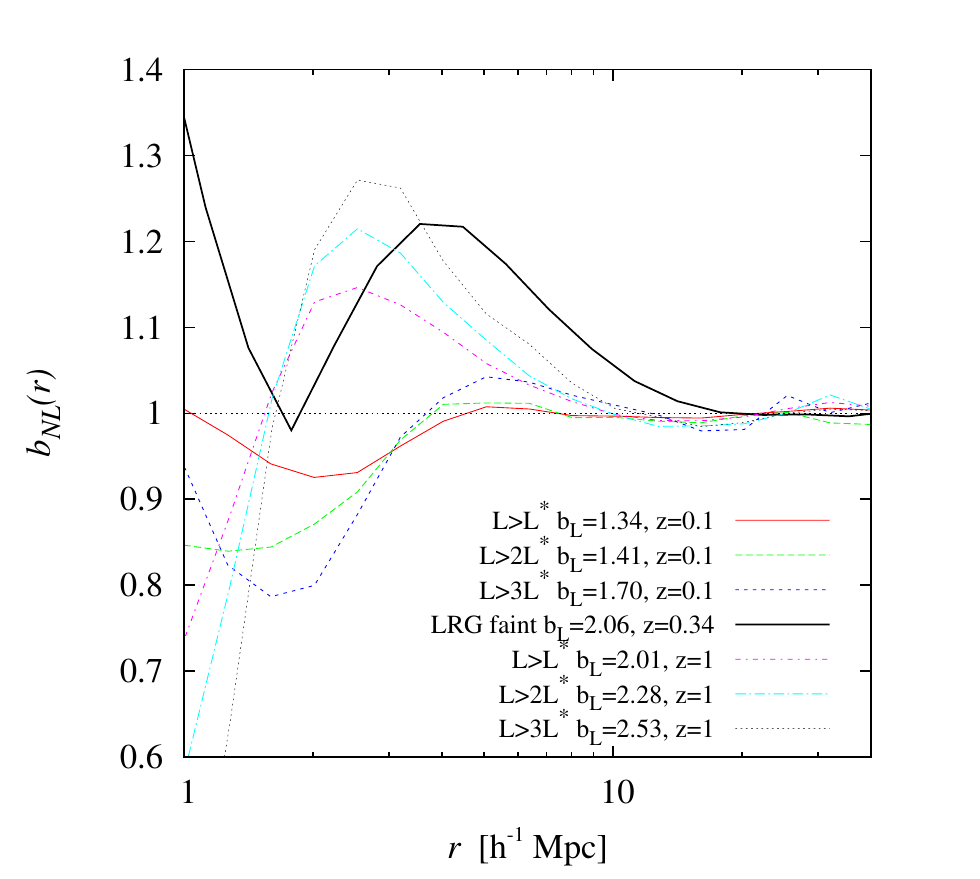}
\caption{The scale dependence of galaxy bias at $z=0.1$ and $z=1.0$,
  for the different galaxy populations considered in this work (see
  inset). It is defined as
  $b_{NL}(r)=\left[\xi_{gg}(r)/\left(b^2_L\xidd(r)\right)\right]^{1/2}$.}
\label{sdbias}
\end{figure}

\begin{figure}
\includegraphics[width=84mm]{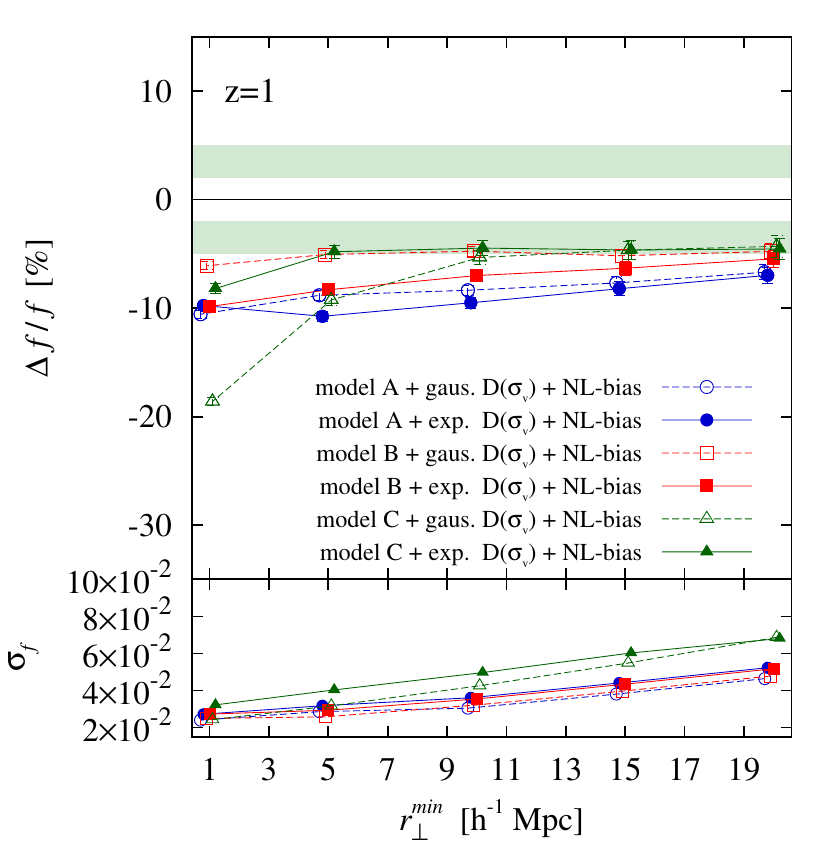}
\caption{Top: relative systematic error on $f$ and its
  corresponding 1$\sigma$ uncertainty for $L>L^*$ galaxies at $z=1$,
  when the bias scale dependence is included. The light (dark) shaded
  band marks the $2\%$ ($5\%$) region around the fiducial
  value. Bottom: 1$\sigma$ statistical error on $f$ in the case of a
  survey probing $1~{\rm h}^{-3}~{\rm Gpc}^3$.}
\label{systbz1}
\end{figure}

\begin{figure}
\includegraphics[width=84mm]{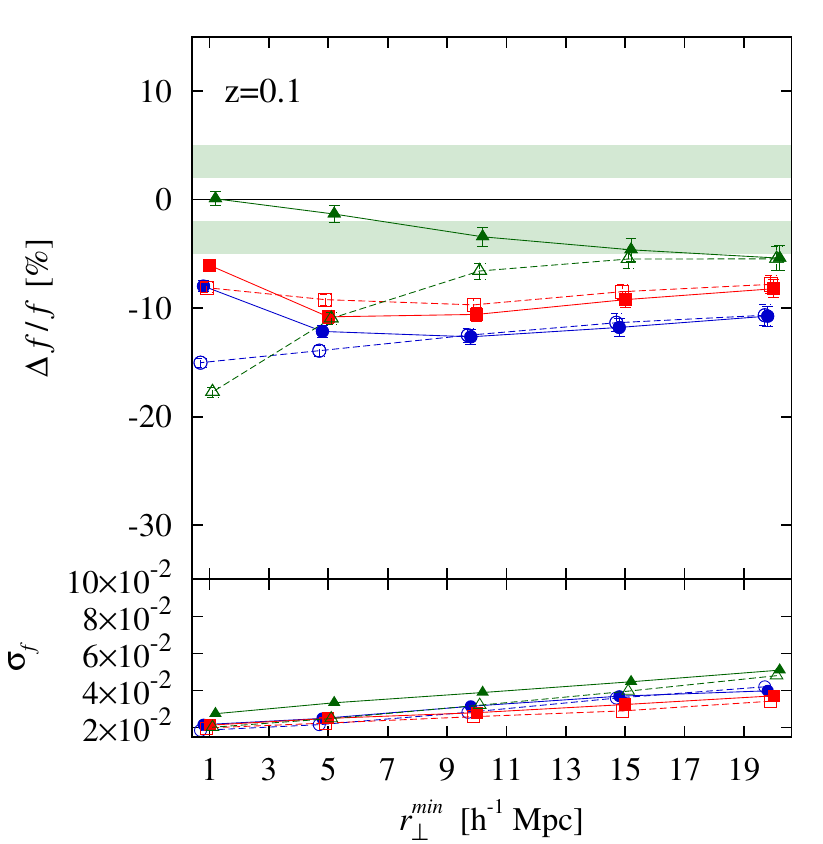}
\caption{Same as Fig. \ref{systbz1} but at $z=0.1$.}
\label{systbz0}
\end{figure}

We now allow for scale-dependence in the galaxy bias description
inside the models and study whether this can improve the recovery of
the growth rate parameter, in particular when including scales below
$10\mhmpc$ in the fitting. In general, the galaxy bias in
configuration space can be defined as,
\begin{equation}
b(r)=\left(\frac{\xi_{gg}(r)}{\xidd(r)}\right)^{1/2}=b_Lb_{NL}(r) \label{eqsdb}
\end{equation}
where $\xi_{gg}$ is the galaxy real-space auto-correlation function
and $b_{NL}(r)$ is the non-linear scale-dependent part of the bias. It
is important to stress that $\xi_{gg}(r)$ is directly measurable from
observations by deprojecting the observed projected correlation
function $w(\rpe)$ \citep{saunders92}. This procedure allows one to
correctly recover the shape of $\xi_{gg}(r)$ up to about $30\mhmpc$
\citep[e.g.][]{saunders92,cabre09a} while it can possibly introduce
noise. In general the latter can increase the statistical error but
may not introduce any systematic bias in the recovery of $f$
\citep{marulli12}, although this has to be investigated in more
details in practical applications. In the following we will therefore
make the assumption that the real-space galaxy auto-correlation
function $\xi_{gg}(r)$ is known, and used its measured values from the
simulated catalogues to infer $b_{NL}(r)$ in the models.  In fact, it
is not necessary to know the exact shape of $\xi_{gg}(r)$ on scales
larger than about $20-30$\hmpc, where one generally finds the galaxy
bias to be almost scale-independent and can thus safely assume
$b_{NL}(r)=1$.  A notable exception is that of more non-linear
objects, for which the scale dependence may extend to larger scales
(see section~\ref{sec:highbias}).

Fig. \ref{sdbias} shows the non-linear scale-dependent component of
galaxy bias, $b_{NL}(r)$, for the different galaxy populations in our
simulated catalogues at the two reference redshifts considered, $z=1$
and $z=0.1$. In the previous section we considered only catalogues of
galaxies with $L>L^*$, while in this figure we introduce more extreme
galaxy populations, which we analyse in the following section. To
define $b_{NL}(r)$, the linear bias $b_L$ has been determined for each
galaxy population by minimising the difference between $\xi_{gg}$ and
$b^2_L\xidd$ on scales above $r=10\mhmpc$. It is evident from this
figure that non-linearities in the galaxy bias produce variations up
to $40\%$ in the real-space clustering on scales $1\mhmpc<r<20\mhmpc$,
the strength of the effect increasing for more luminous galaxies.

Let us come back to our original $L>L^*$ catalogues and repeat the
analysis of the previous section now including the scale dependence of
galaxy bias shown in Fig.~\ref{sdbias}. The new statistical and
systematic errors on $f$ estimated from our simulated catalogues are
shown in Figs. \ref{systbz1} and \ref{systbz0}. In general, one sees
that including the bias scale-dependence information has only the
effect of shifting the recovered $f$ values by about $-3\%$ at both
$z=1$ and $z=0.1$. This systematic effect is not straightforward to
explain but could be due to degeneracies in the models when including
this extra degree of freedom. Accounting for bias scale dependence
tends however to reduce the dependence of the systematic error on the
minimum fitted scale when including scales below $\rpe=10\mhmpc$: the
retrieved value is more constant down to $\rpe^{min}=1\mhmpc$ for all
considered models. Moreover, it reduces the statistical error on $f$
by about $15\%$ for all models.

These results suggest that including the bias scale-dependence
empirically in the models in the way presented here, does not
significantly improve the modelling of $\xisp$ on scales below
$10\mhmpc$. A more detailed inclusion of galaxy bias and its
non-linearities in redshift-space distortions models might be needed.

\subsubsection{Fidelity in reproducing the anisotropic two-point correlation function}

We visually compare in Fig. \ref{contours_z1} and Fig.
\ref{contours_z0} the model correlation functions at $z=1$ and $z=0.1$
using (a) the fiducial values of the parameters
($f$,~$\sigma_v$,~$b_L$) and (b) their best-fit values obtained with
the chosen model, to those measured in our simulated catalogues. For
case (a), $\sigma_v$ is set to linear theory prediction which is given
by Eq. \ref{sigmav}, replacing $\Ptt$ by the linear power spectrum.
We limit this comparison to the case with exponential damping,
scale-dependent bias, and $\rpe^{min}=5\mhmpc$.

As shown in the top panels, when all parameters are fixed to their
fiducial value, model C (solid line) provides a good description of
the observed $\xisp$ in the catalogues, with a slightly worse match
for $\rpe<10\mhmpc$, $\rpa>10\mhmpc$ particularly at $z=1$.
Conversely, model A and B produce contours that at intermediate
separations ($\rpe<40\mhmpc$, $\rpa<25-30\mhmpc$) are less squashed
along the line-of-sight than the data. It is important to keep in mind
that our description of model B and C on scales below
$r\simeq10.5~\mhmpc$ may be biased as we lack information on the
small-scale amplitude of $\xidt(r)$ and $\xitt(r)$ in their
construction (see Section \ref{sec:models}). When
($f$,~$\sigma_v$,~$b_L$) are allowed to vary (lower panels), all
models are generally capable to achieve a good fit to $\xisp$ above
$\rpe=10\mhmpc$.  For model A and B, this is obtained through a lower
damping than predicted by linear theory. This allows one to reproduce
the significant squashing along the line-of-sight seen in the
catalogues, but fails to model the FoG elongation at smaller $\rpe$.

\begin{figure}
\includegraphics[width=80mm]{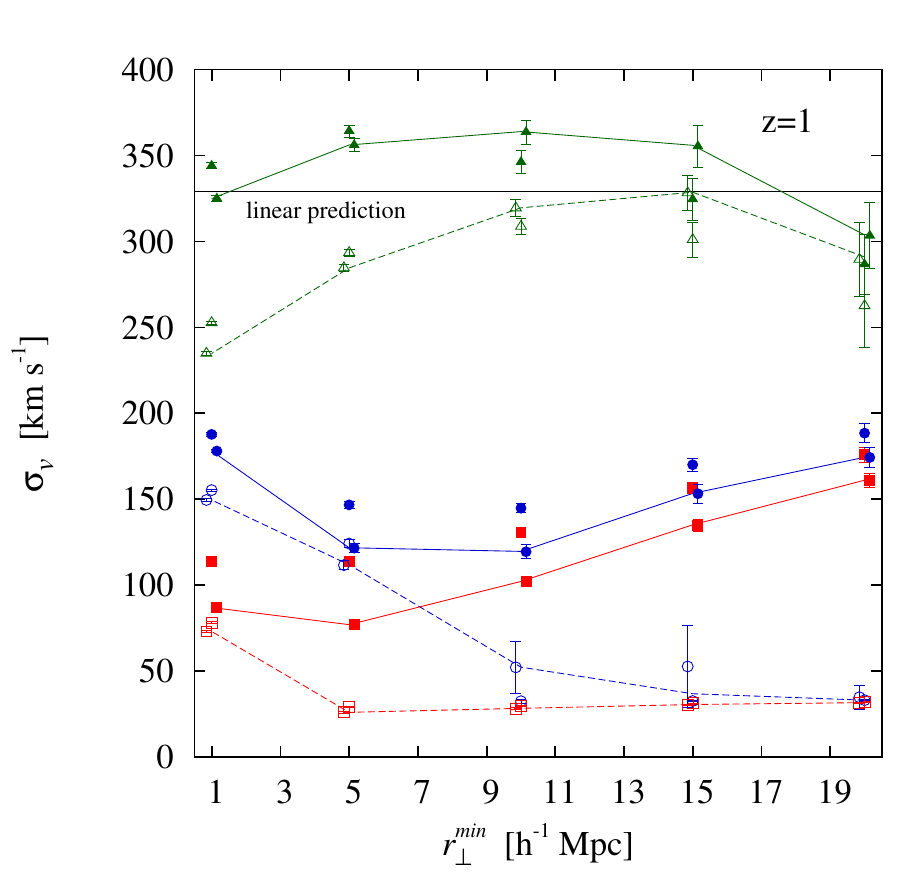}
\caption{Recovered values of $\sigma_v$ for the different models in
  the case of $L>L^*$ galaxies at $z=1$. The symbol definition is the
  same as in Fig. \ref{systz1}. The connected points correspond to the
  case with scale-dependent bias, while others correspond to that with
  linear bias. The latter points are slightly shifted horizontally for
  clarity.}
\label{sigvrec_z1}
\end{figure}

\begin{figure}
\includegraphics[width=80mm]{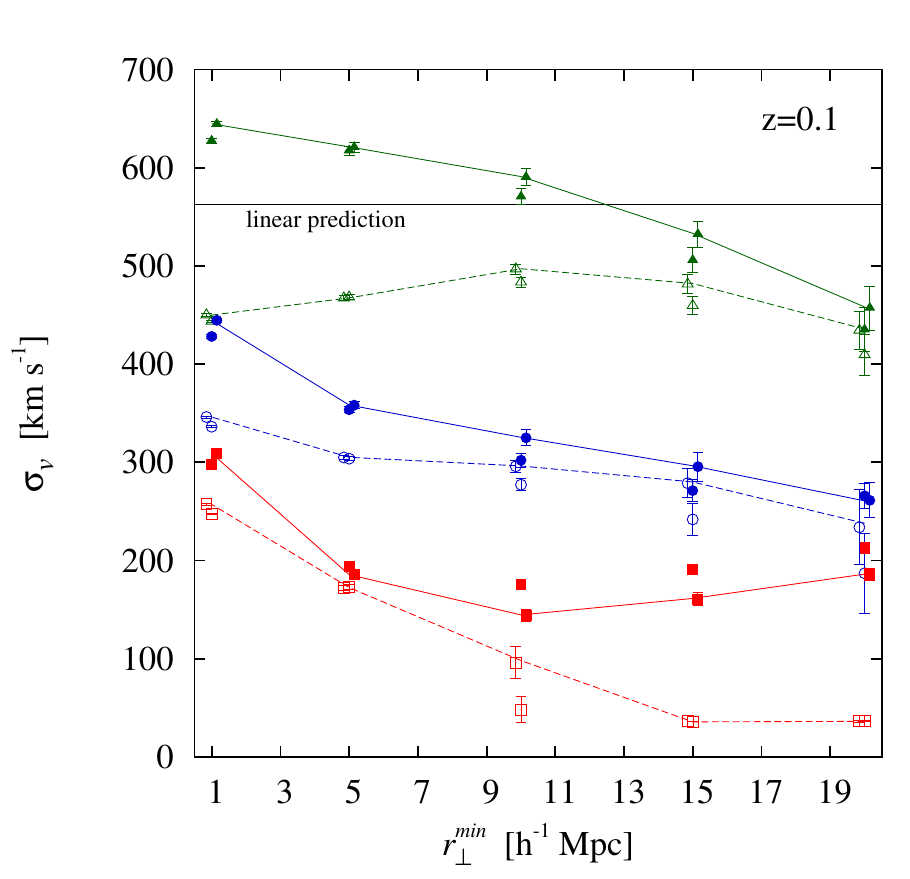}
\caption{Same as Fig. \ref{sigvrec_z1} but at $z=0.1$.}
\label{sigvrec_z0}
\end{figure}

In general, low values of $\sigma_v$ can balance the deficit of power
on small scales, yet they do not allow to recover the true value of
$f$. This is shown in Fig. \ref{sigvrec_z1} and \ref{sigvrec_z0},
where the recovered values of $\sigma_v$ for the different models are
compared to linear prediction. While model C is able to recover
realistic values of $\sigma_v$, of the order of linear theory
predictions, model A and B provides best-fitting values strongly
deviating from linear expectations, leading to $\sigma_v$ as small as
$0-200 \mkms$. These results are consistent with the findings of
\citet{taruya10} and \citet{nishimichi11}, who compared model A and C
in the case of dark matter and halo catalogues, finding a better
agreement of the recovered $\sigma_v$ with linear theory in the case
of model C. These findings confirm that model C is probably less
degenerate than A and B regarding its description of streaming and
random velocities.

\subsubsection{The case of highly-biased galaxies}\label{sec:highbias}

\begin{figure}
\includegraphics[width=84mm]{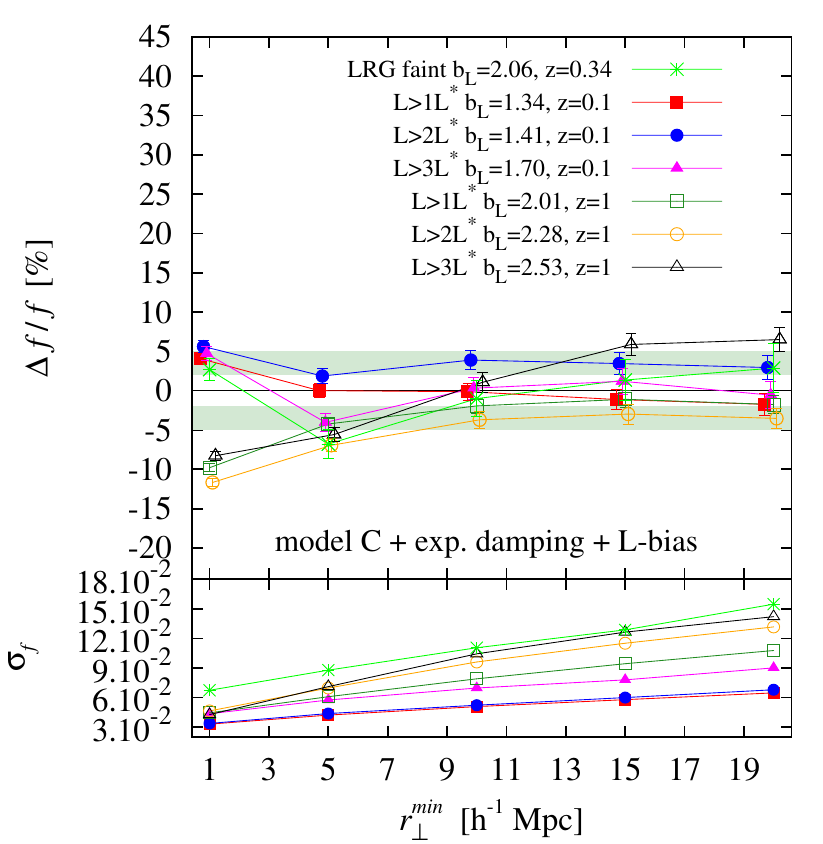}
\caption{Top: relative systematic error on $f$ and its
  corresponding 1$\sigma$ uncertainty for different galaxy populations
  at various redshifts (see inset). In all cases we use model \cexp
  with linear bias. The light (dark) shaded band marks the $2\%$
  ($5\%$) region around the fiducial value. Bottom: 1$\sigma$
  statistical error on $f$ in the case of a survey probing $1~{\rm
    h}^{-3}~{\rm Gpc}^3$, except in the case of the LRG sample where
  the quoted value is for a survey of $13.8~{\rm h}^{-3}~{\rm Gpc}^3$
  (see text for details).}
\label{systall_lin}
\end{figure}

\begin{figure}
\includegraphics[width=84mm]{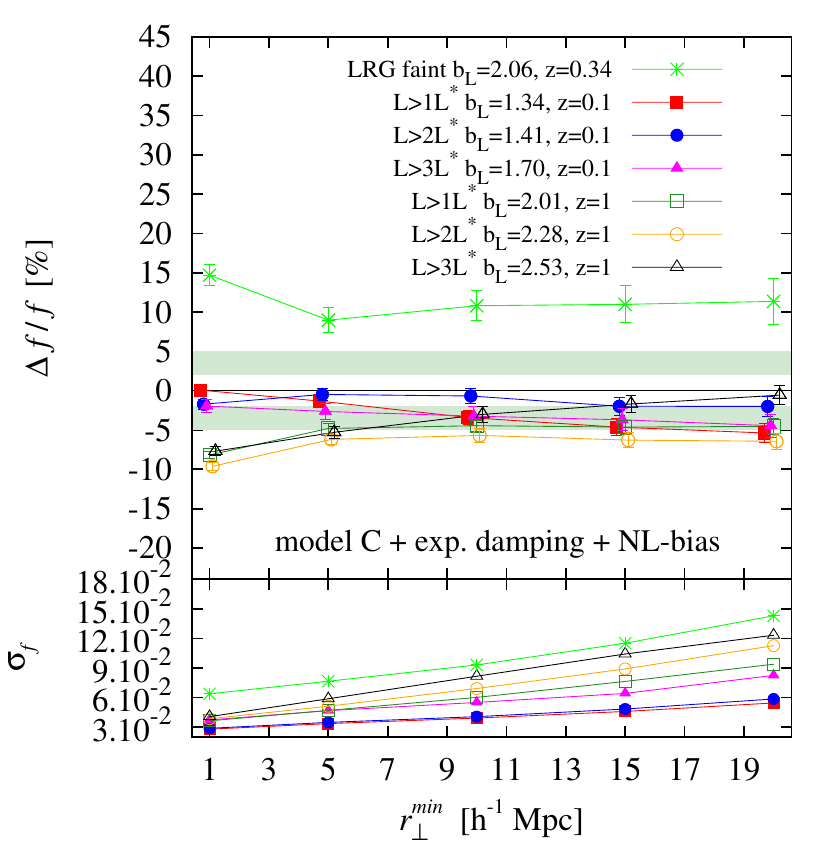}
\caption{Same as Fig. \ref{systall_lin} but in the case where the bias
  is scale-dependent.}
\label{systall}
\end{figure}

Highly biased objects are in general favoured tracers by
redshift-space distortions studies by virtue of the fact that they
probe larger volumes of the Universe.  However, these objects, which
are more likely to reside in the most massive haloes, have undergone a
stronger non-linear evolution, this explaining their stronger bias
scale-dependence seen in Fig. \ref{sdbias}. As such, the inclusion of
$b_{NL}$ in the models may become even more critical for these
objects, if one's goal is to accurately measure the growth rate
parameter \citep[e.g.][]{cabre09b}. In this section, we explicitly
test this hypothesis, extending the model comparison to the regime of
highly biased galaxy populations.

Let us first define higher-luminosity galaxies from our simulated
catalogues. We shall consider two sub-samples, defined as including
galaxies with $L>2L^*$ and $L>3L^*$. We complement these with a
catalogue of simulated Luminous Red Galaxies (LRG) drawn from the
``LasDamas'' suite of simulations, meant to accurately reproduce the
galaxy clustering in the SDSS-DR7 release (McBride et al., in
preparation). More precisely, we shall use 100 mock realisations of
what are defined as ``faint LRG'' ($M_r<-21.2$). These mock samples,
which are in fact lightcones, have been constructed by populating
haloes with galaxies in 40 cubic dark matter N-body simulations of
$2400\mhmpc$ a side and resolution
$m_p=45.73\times10^{10}~\rm{h^{-1}}~\rm{M_\odot}$. When fitting the
measured $\xisp$ in those catalogues, we use in the models the
non-linear mass power spectrum given by CAMB for the cosmological
parameters of the LasDamas simulations. This includes non-linear
evolution of clustering as described by \citet{smith03}.

We compare in Fig. \ref{systall_lin} and Fig. \ref{systall} the
relative systematic error on the growth rate obtained for these highly
biased galaxies. We use here model \cexp with and without including
the scale-dependence of bias. In the case of the linearly-biased
model, the systematic error on the growth rate remains within $\pm
5\%$ at $\rpe^{min}>10\mhmpc$, for all considered galaxy populations,
but those with $L>3L^*$ at $z=1$, which are the most biased objects
considered in this analysis. In the latter case the growth rate is
overestimated by about $6-7\%$, suggesting additional non-linear
effects that are not accounted for in our models.

When including scale-dependent bias, for all but LRG, the dispersion
among the recovered growth rates is reduced and systematic errors
remain within $-7\%<\Delta f/f<-2\%$ when $\rpe^{min}>5\mhmpc$. In the
case of LRG, the growth rate is overestimated by about $10\%$ and the
associated statistical error is higher. It is important to mention
that LRG statistical errors on $f$ in the figures correspond to a
cosmological volume larger by $2.4^3$ than for the other
samples. Thus, in order to make a fair comparison, one has to further
multiply the quoted LRG statistical errors by the square root of the
ratio between the volumes \citep[e.g.][]{bianchi12}, i.e. by about
$3.7$. A cautionary remark in interpreting this result is that the
simulated LRG samples are, unlike the other samples considered,
relatively wide lightcones and as a consequence, may include
wide-angle effects that are not accounted for in our models \citep[but
  see][]{samushia11b}.Moreover, although this sample has been already
used for other investigations, we have no way to verify the details of
the HOD implementation and its impact on our results.

Overall these results confirm our previous findings, suggesting that
for highly non-linearly biased galaxies, additional systematic effects
arise which could be due to an incorrect inclusion of scale-dependence
of bias into the model.  From a theoretical point-of-view, bias
non-linearity changes the correction terms $C_A$ and $C_B$ in model C,
which might need to be modified to properly include galaxy bias
non-linearities and scale-dependence
\citep[see][]{tang11,nishimichi11}. We plan to investigate in more
details these aspects in a future paper. Finally, we note that FoG
modelling could be potentially improved by adding more freedom in the
damping function \citep[e.g.][]{kwan12} or including scale dependence
in the pairwise velocity dispersion \citep[e.g.][]{hawkins03}, but at
the price of increasing the statistical error on the growth rate
estimate.

\section{Effect of galaxy velocity bias} \label{secgalvelbias}

In the framework of understanding the impact of non-linear effects on
the accuracy of growth rate estimates from redshift-space distortions,
we investigate in this section the possible impact of {\it velocity
  bias}, an effect which is is usually neglected.  The galaxy
catalogues that we used so far, assume that the radial distribution of
satellite galaxies inside dark-matter halos follows that of mass, as
described by a \citet{navarro96} (NFW) radial density
profile. Moreover, central galaxies have been defined as being at rest
at the centre of their dark matter halo, inheriting its mean
velocity. These assumptions make galaxy velocities unbiased with
respect to the mass velocity field. However, there are some
observational evidences that galaxies does not exactly follow the same
radial distribution as dark matter and exhibit some velocity bias. In
particular, recent small-scale clustering measurements in the SDSS
suggest that relatively luminous galaxies have a steeper radial
density profile than NFW, with inner slopes close to $-2$ and lower
concentration parameters \citep{watson12}. In addition, galaxy groups
and galaxy clusters analysis tend to indicate that central galaxies
might in general not be exactly at rest at the centre of the potential
well of dark matter haloes
\citep[e.g.][]{vandenbosch05,skibba11}. Differences between the radial
distribution of galaxies and that of mass implies the presence of
additional spatial and velocity biases.  This has a direct impact on
the description of the observed redshift-space distortions.  In this
section we provide a first quantitative assessment of the systematic
uncertainty on $f$ that it can introduce if not accounted for in the
models.

To this end, we include in the catalogues some amount of velocity bias
coming from either central or satellite galaxies. For central galaxies
we follow \citet{vandenbosch05} and assign halo-centric positions
assuming a radial density distribution of the form,
\begin{equation}
\rho_{cen}(r|m)\propto\frac{\displaystyle f_rr_v(m)}{\displaystyle
  r\left(r+f_rr_v(m)\right)^3}
\end{equation}
where $m$ is the halo mass, $r_v$ is the halo virial radius, and $f_r$
is a free parameter that controls the amount of spatial and velocity
bias introduced. This effectively offsets central galaxies from their
halo centre of mass. By solving Jeans equation one obtains the
associated one-dimensional velocity dispersion,
\begin{equation}
\sigma^2_{cen}(r|m)=\frac{1}{\rho_{cen}(r|m)}\int_r^\infty \rho_{cen}(r|m)\frac{d\psi}{dr}dr \label{sigcenloc}
\end{equation}
where $\psi(r)$ is gravitational potential. One can thus define the
velocity bias of central galaxies as,
\begin{equation}
b^{cen}_v(m)=\frac{\left<\sigma_{cen}|m\right>}{\left<\sigma_{dm}|m\right>},
\end{equation}
where the halo-averaged velocity dispersions are given by,
\begin{align}
\left<\sigma_{cen}|m\right>&={}\frac{4\pi}{m}\displaystyle\int_0^{r_v(m)}\rho_{cen}(r|m)\sigma_{cen}(r|m)r^2dr \\
\left<\sigma_{dm}|m\right>&={}\frac{4\pi}{m}\displaystyle\int_0^{r_v(m)}\rho_{NFW}(r|m)\sigma_{NFW}(r|m)r^2dr
\end{align}
and the NFW radial density profile is defined as,
\begin{equation}
\rho_{NFW}(r|m)\propto\left(\frac{c_{dm}(m)r}{r_v(m)}\right)^{-1}\left(1+\frac{c_{dm}(m)r}{r_v(m)}\right)^{-2}
\end{equation}
where $c_{dm}$ is the dark matter concentration parameter (see
Appendix B for its definition). Galaxy halo-centric velocities are
drawn from a Gaussian distribution in each dimension with velocity
dispersion given by Eq. \ref{sigcenloc}. We choose the value of $f_r$
in order to have $b^{cen}_v(m)=0.2$. The latter is an average value
motivated by recent observations \citep[e.g.][]{coziol09,skibba11}.

In the case of satellite galaxies we followed a similar
methodology. Based on the recent results of \citet{watson12} in the
SDSS for $M_r<-20.5$ galaxies, we radially distribute satellite
galaxies as to reproduce a generalised radial density profile of the
form,
\begin{equation}
\rho_{sat}(r|m)\propto\left(\frac{c_{sat}(m)r}{r_v(m)}\right)^{-\gamma}\left(1+\frac{c_{sat}(m)r}{r_v(m)}\right)^{-3+\gamma}
\end{equation}
with $c_{sat}(m)=\frac{c_{dm}(m)}{2}$ and $\gamma=2$. As for central
galaxies, we assign satellite galaxy velocities from the associated
one-dimensional velocity dispersion,
\begin{equation}
\sigma^2_{sat}(r|m)=\frac{1}{\rho_{sat}(r|m)}\int_r^\infty
\rho_{sat}(r|m)\frac{d\psi}{dr}dr \label{sigsatloc}.
\end{equation}
This leads to a velocity bias,
\begin{equation}
b^{sat}_v(m)=\frac{\left<\sigma_{sat}|m\right>}{\left<\sigma_{dm}|m\right>},
\end{equation}
where
\begin{equation}
\left<\sigma_{sat}|m\right>=\frac{4\pi}{m}\displaystyle\int_0^{r_v(m)}\rho_{sat}(r|m)\sigma_{sat}(r|m)r^2dr.
\end{equation}
In that case, $b^{sat}_v(m)$ slowly increases from $b^{sat}_v=1$ at
$m=10^{9}~h^{-1}~M_\odot$ to $b^{sat}_v\simeq1.2$ at
$m=10^{16}~h^{-1}~M_\odot$. Assuming that the haloes are spherical and
isotropic, all previous integrals can be solved analytically
\citep[see e.g. Appendix B and][]{vandenbosch05}.

\begin{figure}
\includegraphics[width=84mm]{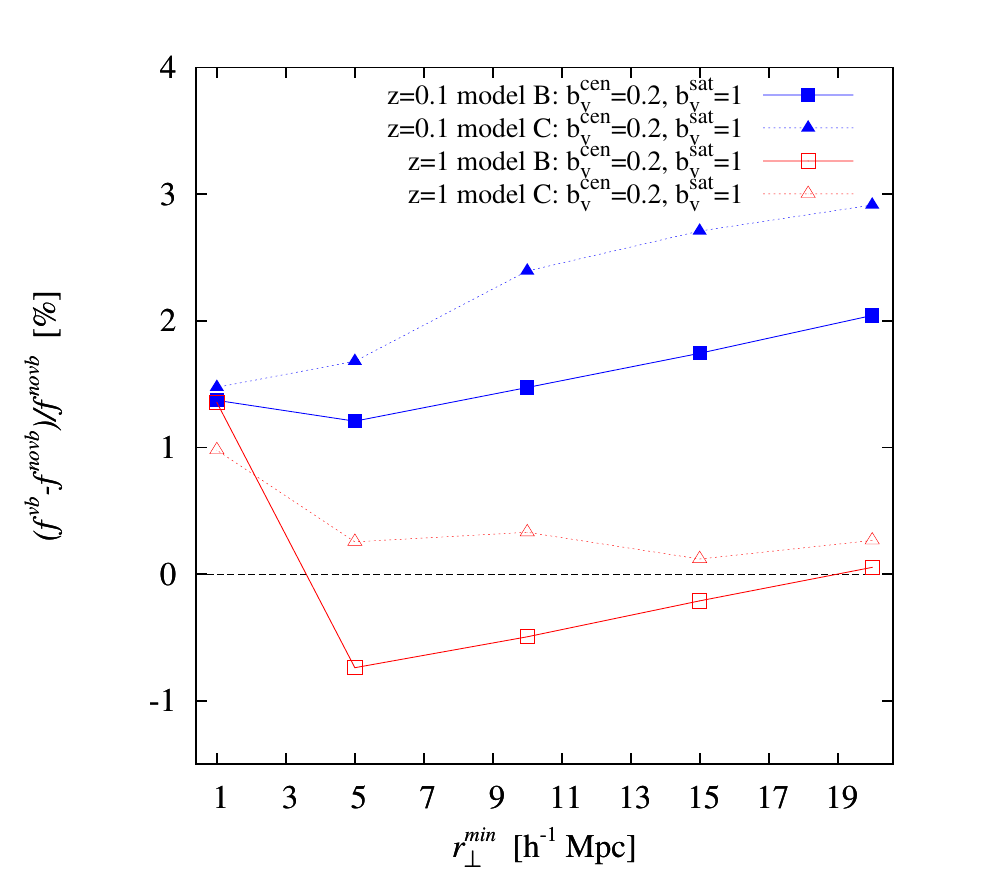}
\caption{Percent systematic variation of the measured growth rate
  $(f^{vb}-f^{novb})/f^{novb}$ induced by including some amount of
  velocity bias from central galaxies in the simulated
  catalogues. Galaxies with $L>L^*$ at the two reference redshifts are
  considered, applying models \bexp and \cexp with scale-dependent
  bias.}
\label{systbv1}
\end{figure}

\begin{figure}
\includegraphics[width=84mm]{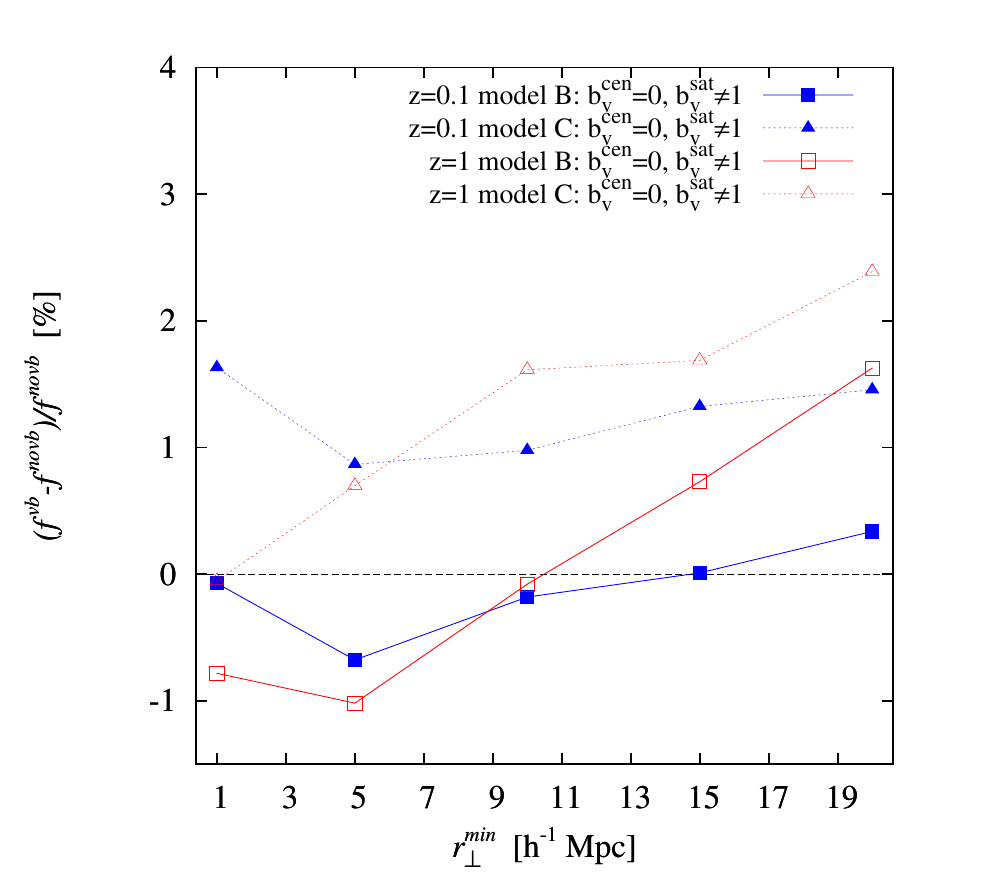}
\caption{Same as Fig. \ref{systbv1} but when including velocity bias
  coming from satellite galaxies in the simulated catalogues.}
\label{systbv2}
\end{figure}

Fig. \ref{systbv1} and \ref{systbv2} show the percent variation on the
estimated $f$, when a velocity bias is included in the simulated
catalogues according to the previously described procedure. The curves
show the systematic effect of velocity bias coming from either central
or satellite galaxies, when $f$ is estimated from models \bexp and
\cexp assuming a scale-dependent spatial bias. We find that the impact
of central galaxy velocity bias is smaller than that of satellite
galaxies: at $z=0.1$ a velocity bias of $b_v^{cen}=0.2$ introduces a
negative systematic error of about $1\%$, independently of the
model. At $z=1$ the effect is larger, with a systematic error that can
reach about $-3\%$ over the range $1\mhmpc<\rpe^{min}<20\mhmpc$.  We
should note that the test at $z=1$ may be pessimistic, as the amount
of velocity bias introduced in the catalogues corresponds to
observational constraints from the local Universe and it is plausible
that at $z=1$ the actual velocity bias is lower. In the case of
satellite galaxies, the introduction of a velocity bias has a stronger
impact: at $z=1$ the systematic error remains within $1-2\%$, while at
$z=0.1$ it is of about $2-3\%$, depending on the model used. In the
latter case, we note that model C tend to be less affected.

These admittedly simple tests suggest that velocity bias, if not
accounted for in the models, can introduce additional systematic
errors of the order of $1-3\%$, i.e. of the same order of statistical
errors expected to be reachable by future large surveys. Although not
dramatic, this additional source of systematic error needs to be kept
in mind and possibly accounted for in future models. In principle,
galaxy velocity bias can be included in the models by adding an
effective velocity bias factor in front of the terms involving the
velocity divergence field. At first approximation, one could assume
this velocity bias factor to be constant and set it as a free
parameter while fitting redshift-space distortions. Introducing more
degrees of freedom in the models would however inevitably increase
statistical errors.

\section{Summary and conclusions}

Measurements of the growth rate of structure from redshift-space
distortions have come to be considered as one of the most promising
probes for future massive redshift surveys that aim at solving the
dilemma of cosmic acceleration. A notable example is the survey
planned for the approved ESA Euclid space mission
\citep{laureijs11}. In this paper, we have investigated in some detail
how well non-linear effects can be accounted for by current models
when performing such measurements on galaxy catalogues.  The question
is how to optimally use observations and models, in order to minimise
statistical and systematic errors alike.  The actual extraction of the
redshift-space distortions signal from real data is subject to two
competing requirements. On one hand, one would like to use the simple
linear description of redshift-space distortions induced by
large-scale coherent motions, thus limiting the measurements and
modelling to very large scales. On such scales fluctuations are close
to be linear and systematic effects appear to be reduced
\citep[e.g.][]{samushia11b}. On the other hand, the clustering signal
on those scales is so weak that statistical errors on the measured
growth rate remain large, even when using samples probing large
volumes. One may therefore prefer to extend the modelling to smaller
scales, enlarging the range of analysed scales and thus reducing the
statistical error.

The most effective compromise to exploit the size and statistics of
future surveys seems thereby that of including intermediate
quasi-nonlinear scales, at the expense of a more complicated modelling
effort. As discussed in the introduction, most recent developments in
this direction have so far concentrated on improving the description
of non-linear effects in the redshift-space clustering of matter
fluctuations in Fourier space. Here we have investigated how these
models perform when applied to catalogues of realistic galaxies in
configuration space, in both the local Universe at $z=0.1$ and the
more distant Universe at $z=1$. For this purpose, we have reformulated
in terms of the anisotropic two-point correlation function of galaxies
$\xisp$, the analytical model for the redshift-space anisotropic power
spectrum by \citet{scoccimarro04} as well as the recent improvements
proposed by \citet{taruya10}, in addition to the standard dispersion
model. In these models, we have included the possibility of using a
realistic non-linear scale-dependent galaxy bias, the latter being in
principle measurable from the observations. At variance with the usual
habit of fitting for the distortion parameter $\beta=f/b_L$, we have
considered the possibility of including the linear component of galaxy
bias as a free parameter and directly estimate the growth rate of
structure $f$. Our key findings and results can be summarised as
follows:
\begin{enumerate}

\item When applied to the galaxy anisotropic correlation function,
  \citet{taruya10}'s model, the most sophisticated model considered in
  this analysis, generally provide the most unbiased estimates of the
  growth rate of structure $f$, retrieving it at the level of about
  $\pm 4\%$ at $z=0.1$ and $z=1$. The commonly used
  \citet{scoccimarro04} and dispersion models generally underestimate
  the growth rate by $4-7\%$ and $5-10\%$ respectively.

\item The inclusion of the scale-dependence of bias in the models is
  important for minimising the systematic error on $f$, in particular
  when one uses the scales below about $10\mhmpc$.

\item Systematic errors vary with the degree of non-linearity in the
  bias of the considered galaxy population, which in turn is a
  function of redshift. Accounting for it is therefore
  particularly relevant when ``slicing'' deep surveys to measure
  $f(z)$ over a significant redshift range.

\item Galaxy velocity bias could represent, if not accounted for, an
  additional source of systematic error.  By implementing realistic
  prescriptions for galaxy velocity bias coming from either central or
  satellite galaxies in our simulated samples, we estimate that
  neglecting it yields to an underestimate of the recovered $f$ by
  $1-3\%$.
\end{enumerate}

Overall, these results further emphasise the need for careful
modelling of non-linear effects, if redshift-space distortions have to
be used as a precision cosmology probe. Further investigation is
needed on the proper inclusion of bias non-linearities, in particular
when using galaxy populations with strong bias scale-dependence. This
is still a non-negligible source a systematic error that has to be
accounted for, to reach the percent accuracy on the growth rate.  Our
results nevertheless indicate promising venues along which to develop
further methods to overcome systematic biases and support the hope
that redshift-space distortions in future surveys will provide us with
a unique test of the cosmological model.

\section*{Acknowledgments}

We thank Atsushi Taruya for providing us with his codes to compute
Closure Theory predictions and for giving us useful comments on the
manuscript. We also thank Davide Bianchi, John Peacock, and Shaun Cole
for discussions and suggestions. Financial support through PRIN-INAF
2007 and 2008 grants and ASI COFIS/WP3110 I/026/07/0 is gratefully
acknowledged.

\bibliographystyle{aa}
\bibliography{biblio}

\appendix

\section{Redshift-space anisotropic two-point correlation function for
  the Taruya, Nishimichi \& Saito (2010) model} 

The redshift-space anisotropic two-point correlation function is
obtainable by Fourier-transforming the anisotropic redshift-space
power spectrum $P^s(k,\mu)$ as,
\begin{equation} 
\xisp = \int \frac{d^3\bmath{k}}{(2\pi)^3} e^{i\bmath{k}\cdot\bmath{s}}
P^s(k,\mu) = \sum_{l} \xi^s_l(s)L_l(\nu)
\end{equation}
where $\nu=\rpa/s$, $\rpe=\sqrt{s^2-\rpa^2}$, and $L_l$ denote Legendre
polynomials. The correlation function multipole moments
$\xi^s_l(s)$ are defined as,
\begin{equation}
\xi^s_l(s)=i^l \int \frac{dk}{2\pi^2} k^2 P^s_l(k)j_l(ks), \label{ximult}
\end{equation} 
where $j_l$ denotes the spherical Bessel functions and
\begin{equation}
P^s_l(k)=\frac{2l+1}{2} \int_{-1}^1 d\mu P^s(k,\mu) L_l(\mu). \label{pkmult}
\end{equation}

In the case of biased tracers of mass, \citet[][]{taruya10} model
for the redshift-space anisotropic power spectrum can be written as,
\begin{align}
P^s(k,\mu)&={} D(k\mu\sigma_v)\left[b^2\Pdd(k)+2b\mu^2 f \Pdt(k) \right. \nonumber \\ 
&\left. + \mu^4 f^2 \Ptt(k) + C_A(k,\mu;f,b) + C_B(k,\mu;f,b)\right] \label{modtargal}
\end{align}
where $b$ is the spatial bias of the considered tracers and,
\begin{align}
C_A(k,\mu;f,b)&={}\sum_{m,n=1}^3 b^{3-n} f^{n} \mu^{2m} P_{Amn}(k), \nonumber \\
C_B(k,\mu;f,b)&={}\sum_{n=1}^4 \sum_{a,b=1}^2 b^{4-a-b} (-f)^{a+b} \mu^{2n} P_{Bnab}(k), \nonumber
\end{align}
with,
\begin{align}
P_{Amn}(k)&={}\frac{k^3}{(2\pi)^2} \left[\int_0^\infty dr \int_{-1}^{+1} dx \left(A_{mn}(r,x)P(k)\right. \right. \nonumber\\
&\left.+ \widetilde{A}_{mn}(r,x) P(kr)\right) \times \frac{P\left(k\sqrt{1+r^2-2rx}\right)}{(1+r^2-2rx)^2} \nonumber \\
&\left.+ P(k) \int_0^\infty dr a_{mn}(r) P(kr) \right],  \label{A_corr}
\end{align}
\begin{align}
P_{Bnab}(k)&={}\frac{k^3}{(2\pi)^2}\int_0^\infty dr \int_{-1}^{+1} dx B^n_{ab}(r,x) \nonumber\\
& \frac{P_{a2}\left(k\sqrt{1+r^2-2rx}\right)P_{b2}(kr)} {(1+r^2-2rx)^a},  \label{B_corr}
\end{align}
where functions $A_{mn}(r,x)$, $\widetilde{A}_{mn}(r,x)$,
$a_{mn}(r,x)$, $B_{ab}(r,x)$ are given in Appendix A of
\citet{taruya10}, $P(k)$ is the linear mass power spectrum,
$P_{12}(k)=\Pdt(k)$, and $P_{22}(k)=\Ptt(k)$.  By using the Kaiser
term in Eq. \ref{modtargal} (i.e. Eq. \ref{modtargal} without the
damping function $D(k\mu\sigma_v)$) into Eq. \ref{pkmult} and
\ref{ximult}, one obtains the corresponding correlation function
multipole moments. The non-null multipole moments are then given by,
\begin{align} 
\xi^{s}_0(s) {} &= b^2\xidd + bf\frac{2}{3}\xidt + f^2\frac{1}{5}\xitt \nonumber \\
&+ b^2f\frac{1}{3}\xi_{A11} + bf^2\frac{1}{3}\xi_{A12} + bf^2\frac{1}{5}\xi_{A22} + f^3\frac{1}{5}\xi_{A23} \nonumber \\
&+ f^3\frac{1}{7}\xi_{A33} + b^2f^2\frac{1}{3}\xi_{B111} - bf^3\frac{1}{3}\left(\xi_{B112}+\xi_{B121}\right) \nonumber \\ 
&+ f^4\frac{1}{3}\xi_{B122} + b^2f^2\frac{1}{5}\xi_{B211} - bf^3\frac{1}{5}\left(\xi_{B212}+\xi_{B221}\right) \nonumber \\ 
&+ f^4\frac{1}{5}\xi_{B222} - bf^3\frac{1}{7}\left(\xi_{B312}+\xi_{B321}\right) + f^4\frac{1}{7}\xi_{B322} \nonumber \\  
&+ f^4\frac{1}{9}\xi_{B422} , \\
\xi^{s}_2(s) {} &= bf\frac{4}{3}\xi^{(2)}_{\delta\theta} + f^2\frac{4}{7}\xi^{(2)}_{\theta\theta} \nonumber \\
&+ b^2f\frac{2}{3}\xi^{(2)}_{A11} + bf^2\frac{2}{3}\xi^{(2)}_{A12} + bf^2\frac{4}{7}\xi^{(2)}_{A22} + f^3\frac{4}{7}\xi^{(2)}_{A23} \nonumber \\
&+ f^3\frac{10}{21}\xi^{(2)}_{A33} + b^2f^2\frac{2}{3}\xi^{(2)}_{B111} - bf^3\frac{2}{3}\left(\xi^{(2)}_{B112}+\xi^{(2)}_{B121}\right) \nonumber \\ 
&+ f^4\frac{2}{3}\xi^{(2)}_{B122} + b^2f^2\frac{4}{7}\xi^{(2)}_{B211} - bf^3\frac{4}{7}\left(\xi^{(2)}_{B212}+\xi^{(2)}_{B221}\right) \nonumber \\ 
&+ f^4\frac{4}{7}\xi^{(2)}_{B222} - bf^3\frac{10}{21}\left(\xi^{(2)}_{B312}+\xi^{(2)}_{B321}\right) + f^4\frac{10}{21}\xi^{(2)}_{B322} \nonumber \\ 
&+ f^4\frac{40}{99}\xi^{(2)}_{B422} , \\
\xi^{s}_4(s) {} &= f^2\frac{8}{35}\xi^{(4)}_{\theta\theta} \nonumber \\
&+ bf^2\frac{8}{35}\xi^{(4)}_{A22} + f^3\frac{8}{35}\xi^{(4)}_{A23} + f^3\frac{24}{77}\xi^{(4)}_{A33} + b^2f^2\frac{8}{35}\xi^{(4)}_{B211} \nonumber \\
&- bf^3\frac{8}{35}\left(\xi^{(4)}_{B212}+\xi^{(4)}_{B221}\right) + f^4\frac{8}{35}\xi^{(4)}_{B222} - bf^3\frac{24}{77}\left(\xi^{(4)}_{B312}\right. \nonumber \\
&+ \left.\xi^{(4)}_{B321}\right) + f^4\frac{24}{77}\xi^{(4)}_{B322} + f^4\frac{48}{143}\xi^{(4)}_{B422}, \\
\xi^{s}_6(s) {} &= f^3\frac{16}{231}\xi^{(6)}_{A33} - bf^3\frac{16}{231}\left(\xi^{(6)}_{B312}+\xi^{(6)}_{B321}\right) + f^4\frac{16}{231}\xi^{(6)}_{B322} \nonumber \\
&+ f^4\frac{64}{495}\xi^{(6)}_{B422}, \\
\xi^{s}_8(s) {} &= f^4\frac{128}{6435}\xi^{(8)}_{B422},
\end{align}
where $\xi_{Amn}$ and $\xi_{Bnab}$ are the Fourier conjugate pairs of
$P_{Amn}$ and $P_{Bnab}$ in Eqs. \ref{A_corr} and \ref{B_corr}, and
$\xi^{(l)}_{X}$ are the correlation function multipole moments
associated with $P_{X}$ as defined in Eq. \ref{ximult}. For orders
$l=2$, $l=4$, $l=6$ and $l=8$, the latter can be conveniently
rewritten as,
\begin{align}
\xi^{(2)}_{X}(r)&={} \xi_{X}(r)-3X_2(r) \\
\xi^{(4)}_{X}(r)&={} \xi_{X}(r)+\frac{15}{2}X_2(r)-\frac{35}{2}X_4(r) \\
\xi^{(6)}_{X}(r)&={} \xi_{X}(r)-\frac{105}{8}X_2(r)+\frac{315}{4}X_4(r) \nonumber \\
&+\frac{693}{8}X_6(r) \\
\xi^{(8)}_{X}(r)&={} \xi_{X}(r)+\frac{315}{16}X_2(r)-\frac{3465}{16}X_4(r) \nonumber \\
&-\frac{9009}{16}X_6 +\frac{6435}{16}X_8(r)
\end{align}
where,
\begin{equation}
X_n(r)=r^{-(n+1)}\int_0^r\xi_{X}(r')r'^ndr'.
\end{equation}
These identities, which orders $l=2$ and $l=4$ were already found by
\citet{hamilton92} and \citet{cole94}, are obtained by using
recurrence relations and integral forms of spherical Bessel functions
\citep[see][for details]{toyoda10}.

Although the model predicts non-null multipole moments of orders $l=6$
and $l=8$, we neglected these terms in this analysis, including only
correlation function multipole moments $\xi^{s}_0(s)$, $\xi^{s}_2(s)$,
and $\xi^{s}_4(s)$ (see main text).

\section{HOD galaxy catalogue construction}

We describe in this appendix the method that we used to create
realistic galaxy catalogues from a large N-body dark matter
simulation, based on the Halo Occupation Distribution (HOD) formalism
\citep[e.g.][]{cooray02}.

We used the dark matter haloes identified using a friends-of-friends
algorithm with linking length of $l=0.17$ in the MDR1 dark matter
simulation by \citet{prada11}. Halo masses were estimated from the sum
of dark matter particle masses inside each halo after correcting for
finite force and mass resolution \citep[][Eq. 4]{bhattacharya11}. We
populated haloes according to their mass by specifying the galaxy halo
occupation which we parametrise as,
\begin{equation}
  \left<N_{gal}|m\right>=\left<N_{cen}|m\right>(1+\left<N_{sat}|m\right>)
\end{equation}
where $\left<N_{cen}|m\right>$ and $\left<N_{sat}|m\right>$ are the
average number of central and satellite galaxies in a halo of mass
$m$. Central and satellite galaxy occupations are defined as
\citep{zheng05},
\begin{align}
  &\left<N_{cen}|m\right>&=&\frac{1}{2}\left[1+\rm{erf}\left(\frac{\log~m - \log
      M_{min}}{\sigma_{\log~m}}\right)\right], \label{ncen} \\
  &\left<N_{sat}|m\right>&=& \left(\frac{m-M_0}{M_1}\right)^{\alpha} \label{nsat}.
\end{align}
where $M_{min}$, $\sigma_{\log~m}$, $M_{0}$, $M_{1}$, and $\alpha$ are
HOD parameters.

We positioned central galaxies at halo centres with probability given
by a Bernoulli distribution function with mean taken from
Eq. \ref{ncen} and assigned host halo mean velocities to them. The
number of satellite galaxies per halo is set to follow a Poisson
distribution with mean given by Eq. \ref{nsat}. We assumed that
satellite galaxies follow the spatial and velocity distribution of
mass and randomly distributed their halo-centric radial position as to
reproduce a \citet{navarro96} (NFW) radial profile,
\begin{equation}
\rho_{NFW}(r|m)\propto\left(\frac{c_{dm}(m)r}{r_v(m)}\right)^{-1}\left(1+\frac{c_{dm}(m)r}{r_v(m)}\right)^{-2}
\end{equation}
where $c_{dm}$ is the concentration parameter and $r_v(m)$ is the
virial radius defined as,
\begin{equation}
r_{v}(m)=\left(\frac{3m}{4\pi\bar{\rho}(z)\Delta_{NL}}\right)^{1/3}.
\end{equation}
In this equation, $\bar{\rho}(z)$ is the mean matter density at redshift $z$
and $\Delta_{NL}=200$ is the critical overdensity for virialisation in
our definition. Consistently with the works of \citet{zehavi11},
\citet{zheng07}, and \citet{coupon12} that we used to infer the mean
galaxy occupation, we assumed the mass-concentration relation of
\citet{bullock01},
\begin{equation}
c_{dm}(m,z)=\frac{c_0}{1+z}\left(\frac{m}{m_*}\right)^{\beta}
\end{equation}
where $c_0=11$, $\beta=-0.13$, and $m_*$ is the non-linear mass scale
at $z=0$ defined such as $\sigma(m_*,0)=\delta_c$. Here $\delta_c$ and
$\sigma(m,0)$ are respectively the critical overdensity (we fixed
$\delta_c=1.686$) and the standard deviation of mass fluctuations at
$z=0$. The latter is defined as,
\begin{equation}
  \sigma^2(m,z)=\int_0^\infty \frac{dk}{k}
  \frac{k^3P(k,z)}{2\pi^2}|W(kR)|^2 \,\,\,\, .
\end{equation}
where $R=\left[3m/\left(4\pi\bar{\rho}(z)\right)\right]^{1/3}$,
$P(k,z)$ is the linear mass power spectrum at redshift $z$ in the
adopted cosmology, and $W(x)$ is the Fourier transform of a top-hat
filter.

In order to assign satellite galaxy velocities, we assumed halo
isotropy and sphericity, and drawn velocities from Gaussian
distribution functions along each Cartesian dimension with velocity
dispersion given by \citep{vandenbosch04},
\begin{align}
\sigma^2_{sat}(r|m)&={}\frac{1}{\rho_{NFW}(r|m)}\int_r^\infty \rho_{NFW}(r|m)\frac{d\psi}{dr}dr \\ 
&={} \frac{Gm}{r_{v}}\frac{c_{dm}}{f(c_{dm})}\left(\frac{c_{dm} r}{r_v}\right)\left(1+\frac{c_{dm} r}{r_v}\right)^2 I(r/rs)
\end{align}
where $\psi(r)$ is the gravitational potential, $G$ is
the gravitational constant, $f(x)=\ln(1+x)-x/(1+x)$, and
\begin{equation}
  I(x)=\int_x^\infty \frac{f(t)dt}{t^3(1+t)^2}.
\end{equation}

\begin{table*}
  \begin{minipage}{150mm}
  \caption{HOD parameters and associated galaxy catalogue properties.}
  \begin{tabular}{@{}ccccccccc}
  \hline
  Redshift & Luminosity &  Absolute magnitude &&&&&& Galaxy number \\
  $z$ & threshold & threshold & $\log M_{min}$ & $\sigma_{\log~m}$ & $\log M_{0}$ & $\log M_{1}$ & $\alpha$ &  density \\
  \hline
  $0.1$ & $L>L^*$  & $M_r-5\log(h)<-20.44$ & 12.18 & 0.21 & 12.18 & 13.31 & 1.08 & 0.347 \\
  $0.1$ & $L>2L^*$ & $M_r-5\log(h)<-21.19$ & 12.89 & 0.68 & 12.89 & 13.89 & 1.17 & 0.105 \\
  $0.1$ & $L>3L^*$ & $M_r-5\log(h)<-21.63$ & 13.48 & 0.70 & 13.48 & 14.35 & 1.23 & 0.026 \\
  \hline
  $1$   & $L>L^*$  & $M_{B/g'}-5\log(h)<-20.78$ & 12.29 & 0.35 & 12.29 & 13.36 & 1.23 & 0.200 \\
  $1$   & $L>2L^*$ & $M_{B/g'}-5\log(h)<-21.53$ & 12.67 & 0.38 & 12.67 & 13.77 & 1.42 & 0.072 \\ 
  $1$   & $L>3L^*$ & $M_{B/g'}-5\log(h)<-21.97$ & 12.94 & 0.40 & 12.94 & 14.10 & 1.54 & 0.032 \\
  \hline
  \label{tab}
  \end{tabular}
  Masses are given in $\rm{h^{-1}}~\rm{M_\odot}$ and galaxy number densities in
  $10^{-2}~\rm{h^{3}}~\rm{Mpc^{-3}}$.
  \end{minipage}
\end{table*}

We generated catalogues corresponding to different populations of
galaxies defined such that their luminosity is greater than multiples
of the characteristic luminosity at both $z=0.1$ and $z=1$. For this
purpose, we used the HOD parameters obtained from the SDSS survey at
$z\simeq0.1$ by \citet{zehavi11} and interpolated the parameter
dependence on luminosity threshold to build catalogues for $L>L^*$,
$L>2L^*$, and $L>3L^*$ galaxies. For $z=1$ catalogues, we used the HOD
parameters measured by \citet{coupon12} and \citet{zheng07}
respectively in the CFHTLS and DEEP2 surveys. In that case, while the
two analysis use slightly different selection magnitude bands ($g'$
and $B$), we assumed here that the later give comparable luminosities
(the two bands largely overlap in wavelength) and interpolate between
the parameters indifferently of the band. The parameter $M_{0}$ is
poorly constrained by current observations and we decided to fix
$M_{0}=M_{min}$. We found that this approximation does not introduce
any significant effect on the halo occupation and predicted
clustering. The characteristic absolute magnitudes at $z=0.1$ and
$z=1$ that we used to define the galaxy samples were taken from
\citet{blanton03} and \citet{ilbert05} respectively. The catalogue
properties and HOD parameters are summarised in Table \ref{tab}.

\bsp

\label{lastpage}

\end{document}